\documentclass[11pt,letterpaper]{article}

\usepackage{amssymb}
\usepackage{amsmath}
\usepackage[bbgreekl]{mathbbol}
\usepackage[scaled=.98,p]{XCharter}
\usepackage[scaled=1.04,varqu,varl]{inconsolata}
\usepackage[type1]{cabin}
\usepackage[uprightscript,xcharter,vvarbb,scaled=1.05]{newtxmath}
\usepackage[cal=euler,scr=boondoxo,bb=boondox]{mathalpha}
\usepackage{anyfontsize}
\usepackage[T1]{fontenc}
\usepackage[final,protrusion=true,expansion=true]{microtype}

\usepackage{titling}
\usepackage{authblk}
\pretitle{\begin{center}\bfseries\sffamily\LARGE}
\posttitle{\end{center}}
\predate{}
\postdate{}

\usepackage{abstract}

\usepackage{sectsty}
\allsectionsfont{\sffamily}

\usepackage{xurl}
\usepackage[left=1in,right=1in,top=1in,bottom=1in,includefoot,heightrounded]{geometry}
\usepackage[margin=15pt,font=small,labelfont={bf,sf},justification=justified]{caption}
\usepackage{graphicx}
\usepackage{flafter}
\usepackage{algorithmic} 
\usepackage{algorithm}
\usepackage{float}
\usepackage{lipsum}
\usepackage{subcaption}
\usepackage{wrapfig}
\usepackage[superscript,biblabel]{cite}
\usepackage{color}
\usepackage{easylist}
\usepackage{lineno}
\usepackage{stmaryrd}
\usepackage{booktabs}
\usepackage[hidelinks]{hyperref}
\usepackage{cleveref}
\usepackage{multirow}
\usepackage{hhline}
\usepackage{soul}
\usepackage{blindtext}
\usepackage{threeparttable}

\def \grad{\nabla}

\def \p   {\partial}
\def \deg {$^{\circ}$}

\renewcommand \d [2]{\frac{{\rm d} #1}{{\rm d} #2}}
\newcommand{\D}[2]{\frac{\p #1}{\p #2}}

\renewcommand{\vec}[1]{\bm{\mathrm{#1}}}

\def \CC{\mathbb{C}}

\def \FF{\mathbb{F}}

\def \PP{\mathbb{P}}

\def \F{\vec{F}}

\def \N{\vec{N}}

\def \X{\vec{X}}

\def \e{\vec{e}}
\def \f{\vec{f}}

\def \n{\vec{n}}

\def \f{\vec{f}}
\def \u{\vec{u}}

\def \x{\vec{x}}
\def \grad{\nabla}

\def \p{\partial}

\def \vchi{\vec{\chi}}

\def \u{\vec{u}}
\def \x{\vec{x}}
\def \tr{\text{tr}}

\def \Da{{\mathrm d} a(\x)}
\def \Dx{{\mathrm d} \x}
\def \DA{{\mathrm d} A(\X)}
\def \DX{{\mathrm d} \X}

\def \Dx{{\mathrm d} \x}

\def \DA{{\mathrm d} A}
\def \Da{{\mathrm d} a}

\def \L2{L^2}

\def \grad{\nabla}

\def \p   {\partial}

\renewcommand{\vec}[1]{\ensuremath\boldsymbol{#1}}

\usepackage[backgroundcolor=white, bordercolor=blue, linecolor=blue, textsize=tiny]{todonotes}

\title{Experimental Bench Validation of a Computational Fluid--Structure Interaction Model of Transcatheter Aortic Valve Dynamics}

\author[1,2,$\dagger$]{Jae H.~Lee} 
\author[3]{Paul Ahern}
\author[3,4]{Anh D. Nguyen\thanks{Present: Harvard John A. Paulson School of Engineering and Applied Sciences (SEAS), Harvard University, Cambridge, MA, USA}}
\author[4]{Masod Sadipour}
\author[5]{Robert Hunt}
\author[3,4]{Wyatt Kmetz}
\author[6,7,8,$\ddagger$]{Boyce E.~Griffith} 
\affil[1]{Department of Mechanical and Aerospace Engineering, Old Dominion University, Norfolk, VA, USA}
\affil[2]{Institute for Engineering in Medicine, Health, and Human Performance, Old Dominion University, Norfolk, VA, USA}
\affil[3]{Department of Mathematics, University of North Carolina, Chapel Hill, NC, USA}
\affil[4]{Department of Biomedical Engineering, University of North Carolina, Chapel Hill, NC, USA}
\affil[5]{Department of Physics, University of Chicago, Chicago, IL, USA}
\affil[6]{Departments of Mathematics and Biomedical Engineering, University of North Carolina, Chapel Hill, NC, USA}
\affil[7]{Computational Medicine Program, University of North Carolina, Chapel Hill, NC, USA}
\affil[8]{McAllister Heart Institute, University of North Carolina, Chapel Hill, NC, USA}
\affil[$\dagger$]{\texttt{jhlee@odu.edu}}
\affil[$\ddagger$]{\texttt{boyceg@email.unc.edu}}

\date{}

\begin{document}


\maketitle

\begin{abstract}

\noindent\textbf{Purpose:}
This study extends our fluid--structure interaction (FSI) framework for surgical bioprosthetic heart valves in an experimental pulse duplicator to a transcatheter aortic valve replacement (TAVR) device.
We assess hemodynamic and leaflet kinematic agreement across pulse rates and the transferability of calibrated boundary-model parameters.

\noindent\textbf{Methods:}
We compared simulated and experimental pressure, volumetric flow rate, projected dynamic valve area (PDVA), and leaflet kinematics at 60, 70, and 80~beats per minute (bpm).
Reduced-order boundary-model parameters were calibrated using only data acquired at 70~bpm and were subsequently held fixed.
At each pulse rate, a condition-specific pump-pressure waveform was constructed from the measured upstream pressure and flow rate.
The 60 and 80~bpm cases therefore evaluate parameter transfer under measurement-informed upstream forcing.

\noindent\textbf{Results:}
Across the three conditions, normalized root-mean-square errors are 2.88--3.08\% for aortic flow rate, 7.64--8.07\% for upstream ventricular pressure, 3.50--6.21\% for downstream aortic pressure, and 3.07--4.70\% for PDVA.
The simulations reproduce the principal pressure, flow, valve-opening, and valve-closure features.
The downstream aortic pressure and PDVA measurements at 60 and 80~bpm were not used to construct the upstream forcing or recalibrate the model parameters and therefore provide the principal out-of-calibration validation evidence.

\noindent\textbf{Conclusion:}
The results provide conditional experimental bench validation evidence for downstream hemodynamics and valve dynamics under measurement-informed upstream forcing and support bound\-ary-parameter transfer across pulse rates without recalibration.
This work advances an experimentally supported, high-fidelity FSI framework for evaluating TAVR device performance.

\end{abstract}

\begingroup
\par
\endgroup

\newpage

\section{Introduction}
\label{sec:introduction}
Aortic stenosis is one of the most prevalent forms of valvular heart disease and can lead to heart failure and death when left untreated.
Nearly 70,000 aortic valve replacement procedures are performed each year in the U.S. to treat severe aortic valve stenosis~\cite{Clark2012}.
Worldwide, over 290,000 procedures are performed each year~\cite{Yoganathan2004,Dasi2009}, and this number is projected to exceed 850,000/year by 2050~\cite{Dasi2009}.
Transcatheter aortic valve replacement (TAVR) has emerged as a minimally invasive alternative to surgical aortic valve replacement for patients with severe aortic stenosis.
In TAVR, a collapsible bioprosthetic valve mounted within a metallic stent is delivered through a catheter and deployed inside the diseased native valve.
The use of bioprosthetic leaflets produces physiological pulsatile valve motion and generally avoids the need for the lifelong anticoagulation required by mechanical valves~\cite{Rodriguez-Gabella2017}.
Despite the clinical success of TAVR, device performance can be affected by complications such as paravalvular regurgitation, leaflet thrombosis, reduced leaflet mobility, structural valve deterioration, and abnormal leaflet stresses~\cite{Dasi2017,Koren2023,Patel2023,Ternacle2024}.
These outcomes depend on the interaction among the deployed valve geometry, leaflet mechanics, surrounding blood flow, and patient-specific loading conditions.
However, the coupled mechanisms governing transcatheter valve motion and hemodynamics remain incompletely understood, limiting further improvements in valve design~\cite{Tango2018}.
A comprehensive understanding of the interaction between leaflet kinematics and fluid dynamics is therefore needed to evaluate TAVR performance, identify mechanisms of device dysfunction, and guide the development of safer and more durable transcatheter valves.

In vitro testing in a cardiovascular pulse duplicator provides a controlled approach for investigating the coupled leaflet kinematics and fluid dynamics of transcatheter heart valves.
Particle image velocimetry (PIV) is widely used for this purpose because it provides spatially resolved velocity measurements from which quantities such as Reynolds stresses and turbulent kinetic energy can be estimated~\cite{Toninato2016}.
Previous studies have used PIV to characterize the flow produced by bioprosthetic and transcatheter heart valves and to examine the influence of valve design and operating conditions on downstream hemodynamics~\cite{Lim1997,Wu2019RoundRobin,Hatoum2018,Hatoum2019,Hatoum2021}.
Despite these advantages, PIV requires careful optical calibration, synchronization, flow seeding, and refractive-index matching.
Measurement uncertainty and cycle-to-cycle variability must also be considered when interpreting phase-resolved flow fields.
In addition, optical access may be limited by the valve leaflets, stent, test-section geometry, or other components of the experimental apparatus.
Consequently, PIV may not provide complete flow information throughout the entire three-dimensional domain.

Computational modeling and simulation (CM\&S) complements pulse-duplicator measurements by providing access to three-dimensional velocity and pressure fields, including regions that are difficult or impossible to observe optically.
Once the computational model is established, operating conditions and model parameters can be controlled precisely, and multiple device configurations or loading conditions can be evaluated systematically.
CM\&S also provides access to derived quantities that may be difficult to measure directly, including leaflet stress and strain, transvalvular pressure distributions, and fluid forces acting on the leaflets.
Its spatial and temporal resolution is determined by the computational discretization and can therefore be selected to resolve flow and structural features relevant to the intended analysis.
The credibility of these predictions, however, depends on appropriate verification and validation for the intended context of use~\cite{Pruett2020}.
Direct comparison with controlled in vitro measurements is therefore necessary to quantify agreement, identify important model-form and parameter uncertainties, and establish confidence in the use of the computational framework for evaluating transcatheter heart valve performance.

Consistent with the terminology used in ASME V\&V~40 and FDA guidance for computational modeling of medical devices, we distinguish model calibration from validation~\cite{ASMEVV40,FDACredibility}.
Calibration uses experimental observations to identify model parameters, whereas validation evaluates the finalized model against observations not used for parameter identification or model development.
Agreement at the calibration condition therefore provides calibration evidence but does not, by itself, establish predictive accuracy at other operating conditions.
In the present study, boundary-model parameters were calibrated at 70~bpm and held fixed for the 60 and 80~bpm simulations.
At each pulse rate, measured upstream pressure and flow informed the prescribed pump pressure waveform.
The 60 and 80~bpm comparisons therefore assess parameter transfer under condition-specific, measurement-informed forcing, with downstream aortic pressure and PDVA as the principal validation outputs.
Comparisons at 70~bpm assess agreement at the calibration condition.

Validating a computational model against a pulse duplicator is most informative if both interrogate the same valve under matched conditions.
Lee et al.~\cite{Lee2023} tested seven polymeric valves in a pulse duplicator under ISO 5840 and reproduced them with an arbitrary Lagrangian--Eulerian (ALE)/immersed-solid fluid--structure interaction (FSI) scheme, matching effective orifice area (EOA) values and the inverse leaflet-thickness--EOA relationship, with residual discrepancies traced to simulating only the first cycle and an inviscid-fluid, rigid-conduit idealization.
Borowski et al.~\cite{Borowski2022validation} performed a particularly relevant validation of a TAVR FSI model using high-speed videography and PIV measurements acquired in a pulse duplicator under ISO~5840 conditions.
The simulated geometric orifice area agreed with the experimental measurements over the systolic interval, and qualitative comparisons of the velocity fields across three measurement planes showed similar peak-systolic flow characteristics.
This study demonstrates the value of simultaneously evaluating leaflet kinematics and local fluid dynamics, although its velocity-field comparison was primarily qualitative and focused on peak systole.
Kaiser et al.~\cite{Kaiser2023} extended this from device metrics to the velocity field, using the immersed boundary (IB) method in IBAMR to reproduce four-dimensional flow magnetic resonance imaging (4D-flow MRI) measurements of a bioprosthetic heart valve (BHV) in a pulse duplicator and achieving $L^1$ errors near 30\% on the pointwise velocity field at mid-systole---a more stringent benchmark than integral descriptors alone.
Zingaro et al.~\cite{Zingaro2026} carried the same IB approach with 4D-flow comparison in vivo, validating patient-specific aortic-valve FSI models against MRI velocity fields in five pre-TAVR patients and reproducing the systolic jet, kinetic energy, and vortical structures within the uncertainty of the measurement itself.
They also noted that velocity-derivative quantities such as wall shear stress remain limited by 4D-flow resolution and noise.
The tolerance for such agreement is itself bounded by bench reproducibility: in an ISO round-robin across thirteen laboratories, Wu et al.~\cite{Wu2019RoundRobin} reported coefficients of variation of 8--22\% for EOA and 15--46\% for mean pressure gradient, providing benchmark variability against which any pulse-duplicator validation must be interpreted.
Beyond the aortic position, pulse duplicators are commonly used for stand-alone bench testing of pulmonary and transcatheter valves~\cite{BreitensteinAttach2025,PerezCualtan2025,Zakko2020}, but studies pairing these experiments with fully coupled FSI models of the valve leaflets remain comparatively limited.
Moreover, most existing studies evaluate a single operating condition or emphasize either integral hemodynamic quantities, leaflet kinematics, or spatially resolved flow rather than considering these validation targets together.

A related body of work focuses on structural validation of TAVR deployment rather than pulsatile valve hemodynamics.
Bosmans et al.~\cite{Bosmans2016} developed patient-specific finite-element models of self-expanding TAVR deployment and compared the simulated stent configurations with postoperative computed tomography (CT) measurements from 10 patients.
Grossi et al.~\cite{Grossi2024} calibrated self-expandable TAV stent models against experimental crimping force--diameter curves, then validated patient-specific deployment against postoperative CT, reporting a credibility framework anchored in device geometry rather than hemodynamics.
These studies establish credibility for predicting deployed device geometry and potential paravalvular gaps, but they do not directly validate the coupled pulsatile flow and leaflet dynamics.

In previous work, we developed FSI models of an experimental pulse duplicator to simulate surgical BHV dynamics~\cite{Lee2020pulse,Lee2021flutter}.
The previous study demonstrated that our FSI models yielded excellent agreement with experimental measurements, including bulk flow rates, pressure waveforms, and valve open-area waveforms, as well as reasonable agreement in valve leaflet kinematics.
We also showed that the large-scale flow features are reasonably resolved at practical computational grid resolutions.
We subsequently used the experimental and computational platforms to investigate the effects of valve geometry on systolic leaflet flutter~\cite{Lee2021flutter}.
Both platforms demonstrated that smaller-diameter valves exhibited substantially higher flutter frequencies across a range of operating conditions, and the computational model predicted that flutter frequency also increased with leaflet thickness.
A scaling model rationalized these relationships, demonstrating how validated FSI simulations can complement experiments by isolating the effects of geometric and mechanical parameters on valve dynamics.
These findings also suggest a possible connection between device geometry, leaflet flutter, and long-term bioprosthetic valve durability.

The present study extends this framework from surgical bioprosthetic valves to a TAVR device.
We compared a fully coupled TAVR FSI model with pulse-duplicator measurements of upstream and downstream pressure, volumetric flow rate, projected dynamic valve area (PDVA), and leaflet kinematics under pulsatile operating conditions.
We evaluated the model at pulse rates of 60, 70, and 80~bpm.
We calibrated the reduced-order boundary-model parameters exclusively using the 70~bpm measurements.
Without further parameter adjustment, we compared downstream aortic pressure and PDVA at 60 and 80~bpm under upstream forcing constructed from the corresponding measured pressure and flow waveforms.
Although the 70~bpm measurements were used for boundary-model calibration, agreement between the corresponding fully coupled FSI simulation and experiment is not imposed because the simulated pressure, flow rate, and leaflet motion emerge from interactions among the fluid, deforming leaflets, and calibrated boundary models.
The resulting comparisons provide bench validation evidence supporting the credibility of the FSI framework for reproducing TAVR hemodynamics and leaflet kinematics over the investigated operating conditions.

\section{Materials and Methods}
\label{sec:methods}

\subsection{Experimental setup}
\label{subsec:pulse_duplicator}
\begin{figure}[tt!]
	\begin{center}
		\includegraphics[scale = 0.4,trim={0 100 0 100},clip]{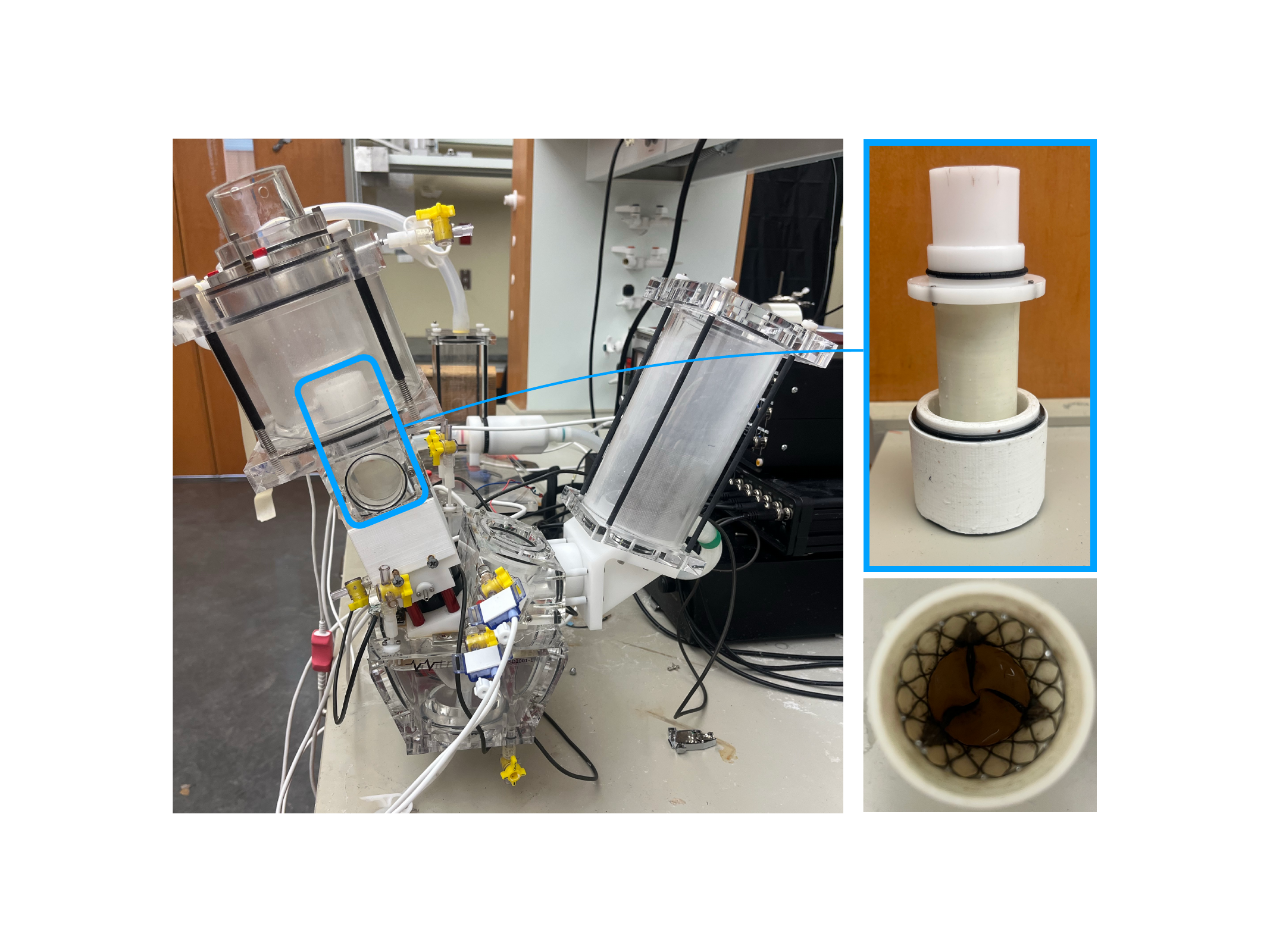}
		\caption{
        Experimental pulse-duplicator system used for in vitro evaluation of the transcatheter aortic valve.
        The left panel shows the complete ViVitro cardiovascular flow-loop apparatus in the Cardiovascular Modeling and Simulation Laboratory at the University of North Carolina at Chapel Hill.
        The outlined region identifies the transparent valve-testing chamber.
        The upper-right panel shows the removable valve holder used to position the transcatheter valve within the test section, and the lower-right panel provides a top view of the implanted valve.
        The transparent test chamber permits high-speed videographic recording of leaflet motion and projected valve opening while pressure and volumetric flow-rate waveforms are measured throughout the cardiac cycle.
        }
		\label{fig:experimental_setup}
	\end{center}
\end{figure}

The experimental studies used a ViVitro Pulse Duplicator (ViVitro Labs, Inc., Victoria, BC, Canada) (Figure~\ref{fig:experimental_setup}).
We used a 26~mm bioprosthetic TAVR valve composed of porcine pericardial tissue (Medtronic CoreValve Evolut R) in our experiments.
This valve was specifically designated for experimental use and stored in glutaraldehyde solution.
Flow and pressure signals were acquired via the pulse duplicator.
Pressure measurements acquired from the pulse duplicator were low-pass filtered at 30~Hz using the Ampack pressure amplifier.
We used saline as the test fluid for this initial study because the viscosities of common blood analogues are more sensitive to temperature than saline.
The ViVitro system can also use high-speed videography to assess valve kinematics or PIV to provide instantaneous and time-averaged flow data.
Saline is widely used for hydrodynamic assessments of heart valves according to the ISO 5840-3 standard~\cite{ISO5840-2021}, which allows saline to be used as a test fluid, and previous studies have done so as well~\cite{Wu2019RoundRobin,Lee2020pulse,Lee2021flutter}.
Our solution also contained 0.9\% NaCl by volume to allow for the bulk-flow measurement using the Carolina Medical flowmeter (Carolina Medical Electronics, Inc., East Band, NC, USA).
The flowmeter was configured to filter the signal at 30~Hz.
Other settings were set to the recommended default values.
The flow probe was mounted upstream and downstream of the aortic site, and the ground electrode was connected to the fluid ground point corresponding to the probe location.
A calibration file was generated using the recommended calibration procedures and instrument settings outlined in the ViVitro Pulse Duplicator System manual.
Three pressure transducers measured pressure at the aortic, ventricular, and atrial positions. The recirculation resistance and outflow compliance volume were adjusted to provide physiological flow conditions.
Flow and pressure waveforms were acquired simultaneously with high-speed videography using a Chronos 2.1-HD camera operating at 500 frames per second.
An LED attached to the aortic chamber observation window and connected to the ViVitro SuperPump’s Sync Out signal synchronized the video recordings with the flow and pressure measurements.
Image contrast was adjusted to improve leaflet visualization, and PDVA was reconstructed from the recordings through automated image analysis in ImageTank (Visual Data Tools, Inc., Chapel Hill, NC, USA).

\subsection{Computational model}
\subsubsection{Porcine pericardial TAVR device geometry}
\label{subsubsec:tavr_geometry}

A three-dimensional volumetric model (Figure~\ref{fig:model_setup}A) was created using Onshape (Onshape Inc., Boston, MA) based on direct measurements.
The inlet and outlet diameters are both 26~mm, corresponding to the inner circular surfaces of the wall annulus on which the upstream and downstream boundary conditions are applied, and the waist diameter is 21~mm.
The stent wires were created by projecting a pattern onto the volumetric solid, then revolving and filleting edges to form the final model.
The porcine pericardial sealing skirt was modeled in SOLIDWORKS (Dassault Systems, Vélizy-Villacoublay, France) by knitting surfaces around the stent frame and extruding the surface to a thickness of 0.34~mm.
The idealized porcine pericardial valve leaflets were generated using MATLAB (MathWorks Inc., Natick, MA) and SOLIDWORKS, with dimensions taken from the physical device (leaflet thickness: 0.34~mm, leaflet diameter: 27~mm, commissure height: 17~mm, free-edge height: 10.625~mm).

\subsubsection{Leaflet biomechanics}
\label{subsubsec:biomechanics}
\begin{figure}[tt!]
	\begin{center}
		\includegraphics[scale = 0.35,trim={150 0 10 0},clip]{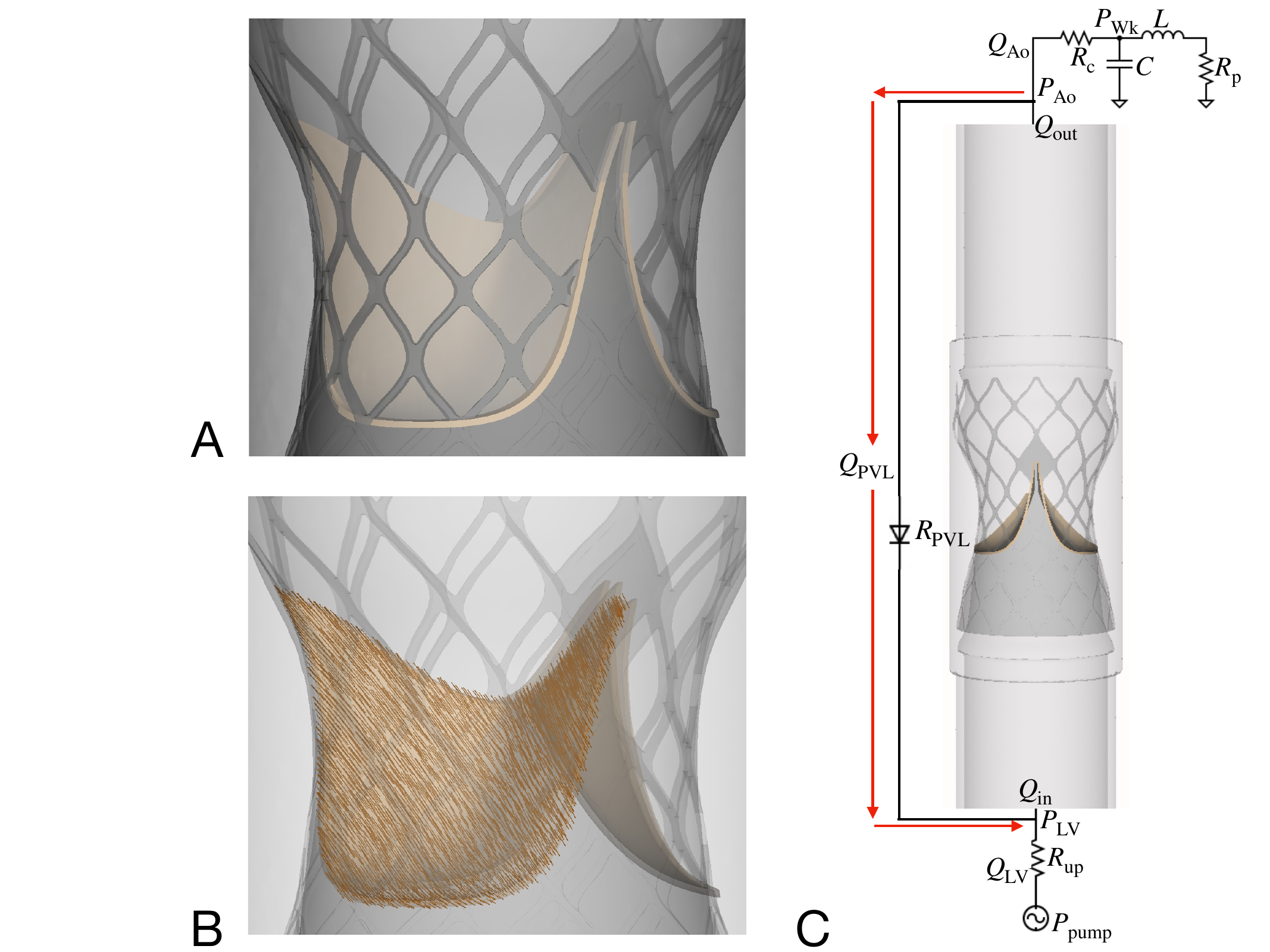}
		\caption{\textbf{A.} Computational model of the porcine pericardial transcatheter aortic valve, including the prosthetic frame and valve leaflets.
        \textbf{B.} Prescribed leaflet anisotropy constructed from small-angle light-scattering measurements reported by Sun et al.~\cite{Sun2005} and following previously described methods~\cite{Lee2020pulse,Lee2021flutter,Brown2023}.
        \textbf{C.} Schematic of the coupled upstream and downstream lumped-parameter boundary models used to impose physiological loading on the computational valve.
        The upstream model relates ventricular pressure, pump pressure, and ventricular inflow, whereas the downstream Windkessel model represents the resistive and compliant loading of the aortic circulation.
        An additional parallel flow path with resistance ($R_\text{PVL}$) represents paravalvular leakage between the upstream and downstream compartments.}
		\label{fig:model_setup}
	\end{center}
\end{figure}

We use a Gasser--Holzapfel--Ogden (GHO) model~\cite{Gasser2006} to describe the biomechanics of the porcine pericardial valve leaflets.
The model is described by the strain-energy functional
\begin{equation}
\Psi(\FF) = C_{10} \left\{\exp\left[C_{01}(\bar{I}_1-3)\right]-1\right\} + \frac{k_1}{2k_2}\sum_{i=1}^{2}\left\{\exp\left[k_2\left(I_{4,i}^{\star}-1\right)^2\right]-1\right\} + U(J),
\end{equation}
in which $\FF$ is the deformation-gradient tensor, $J=\det(\FF)$ is the local volume ratio, $\overline{\FF}=J^{-1/3}\FF$ is the modified, or isochoric, deformation-gradient tensor, $\bar{I}_1=\tr(\overline{\CC})$ is the first invariant of the modified right Cauchy--Green strain tensor $\overline{\CC}=\overline{\FF}^{T}\overline{\FF}$, $I_{4,i}^{\star}=\max(I_{4,i},1)=\max(\e_{0,i}^{T}\CC\e_{0,i},1)$ is the tension-only invariant associated with the $i$th fiber family, and $\e_{0,i}$ is a unit vector aligned with the corresponding fiber direction in the reference configuration.
The term $U(J)$ is the volumetric contribution used to penalize changes in volume.
The mean fiber direction is $45$\deg, and the model includes two fiber families rotated within the leaflet surface by $\pm\theta$ from the mean fiber direction.
The constitutive parameters are fitted to experimental data for glutaraldehyde-fixed porcine pericardial tissue reported by Murdock et al.~\cite{Murdock2018}: $C_{10}=2.32$~kPa, $C_{01}=28.54$, $k_1=0.1097$~MPa, $k_2=91.98$, and $\theta=0.002$\deg.
We also used the previously described induced anisotropy for pericardial BHVs, which is based on experimental measurements of small-angle light-scattering data~\cite{Sun2005} (Figure~\ref{fig:model_setup}B).
This formulation models the valve leaflets as exactly incompressible, but in our numerical formulation, we use the volumetric strain energy $U(J)=\beta\left[J\ln(J)-J+1\right]$ as a stabilization term to reinforce the incompressibility constraint~\cite{Vadala-Roth2020}.

\begin{table}[tt!]
\centering
\caption{Summary of material parameters and their values for the valve leaflets.}
\label{tab:valve_parameters}
\begin{tabular}{llllllll}
\hline
\textbf{Parameter} & \vline & $C_{10}$ & $C_{01}$ & $k_1$ & $k_2$ & $\theta$ & $\beta$ \\
& \vline & [kPa] & --- & [MPa] & --- & [deg.] & [MPa] \\
\hline
\textbf{Value} & \vline & 2.32 & 28.54 & 0.1097 & 91.98 & 0.002 & 5.840\\
\hline
\end{tabular}
\end{table}

\subsubsection{Aortic test-section model}
To reconstruct the geometry of the ViVitro Labs, Inc. aortic test section, a mold of the interior was made using plaster of Paris.
The surface of this mold was scanned using the NextEngine 3D scanner (NextEngine, Inc., Santa Monica, CA, USA).
The TAVR device conduit was created using Onshape with a height of 60~mm, inner top and bottom diameters of 29~mm, a waist diameter of 21.5~mm, and a thickness of 1.5~mm.
The conduit was formed by revolving a sketch that contoured the side profile of the TAVR device.
The conduit was printed on the Bambu X1 (Bambu Lab, Shenzen, China) using a flexible TPU filament (NinjaTek, Lititz, PA).
The inflow and outflow mounts were created using Onshape with the same dimensions as the corresponding original parts from the ViVitro pulse-duplicator system (ViVitro Labs, Victoria, BC, Canada): heights of 49~mm for the top mount and 31~mm for the bottom mount.
The portions of the mounts inserted into the conduit have a thickness of 1.5~mm, an outer diameter of 29~mm (matching the conduit's inner diameter), and a height of 5~mm.
The mounts were printed on the UltiMaker S5 (UltiMaker, Utrecht, Netherlands) using PET-G filament.
For the computational mesh, the mounts' outer details were stripped, turning them into smooth tubes with a thickness of 1.5~mm and an outer diameter of 29~mm.
The tubes were inserted 5~mm into the conduit.

We use a penalty method to model the aortic test section as an essentially rigid material model via
\begin{align}
	W_\text{wall} &= \frac{c_\text{wall}}{2}(\bar{I}_1 - 3), \text{ and}\\
	\F(\X,t) &= \kappa_\text{wall}(\X - \vchi(\X,t)),
\end{align}
in which $c_\text{wall}=33.1~\text{kPa}$ and $\kappa_\text{wall}=8.52\times10^5~\text{kPa}\,\text{cm}^{-2}$.

\subsubsection{Fluid model and boundary conditions}
We use the incompressible Navier--Stokes equations to model the fluid in the aortic mount as a viscous incompressible fluid.
We model saline as a Newtonian fluid with a uniform density $\rho~=~1.01$~g~cm$^{-3}$ and a uniform viscosity $\mu~=~1.0$~cP.

A straightforward approach for the upstream flow-source boundary would be to prescribe the experimentally measured flow-rate waveform directly.
However, doing so would constrain the natural coupling among ventricular pressure, transvalvular flow, and leaflet motion, preventing the valve from experiencing a physiologically realistic, dynamically evolving transvalvular pressure gradient.
Therefore, we use a single-resistance model at the upstream boundary as the simplest representation capable of reproducing the measured flow while preserving realistic pressure--flow coupling.
Because the downstream boundary acts as neither a prescribed flow source nor a prescribed pressure source, we represent it using a Windkessel model.
Our experimental measurements also exhibited inertial behavior, so rather than using the conventional RCR outlet model, we use an RCLR model that includes an additional inertance to capture this effect.

The experimentally measured flow-rate waveform also exhibited paravalvular regurgitation.
The computational geometry included an idealized stent frame in its prescribed deployed configuration, and the model did not simulate stent crimping, deployment, contact, or apposition against the surrounding test-section wall.
Instead, the interface between the implanted valve assembly and the wall was treated as completely sealed, so the resolved fluid could pass only through the valve orifice. 
Consequently, the model could not resolve the nonuniform periprosthetic gaps responsible for the experimentally observed paravalvular leakage.
Because modeling the deployment and contact mechanics required to predict these gaps was beyond the scope of this study, paravalvular leakage was represented by an additional lumped-parameter flow pathway between the downstream and upstream boundary models (Figure~\ref{fig:model_setup}C).
This pathway contains a resistance, $R_\text{PVL}$, and a diode-like element that permits retrograde flow if the downstream aortic pressure exceeds the upstream ventricular pressure.
The resulting leakage flow, $Q_\text{PVL}$, is
\begin{equation}
    \label{eq:pvl_eqn}
    Q_\text{PVL} = \frac{P_\text{LV}-P_\text{Ao}}{R_\text{PVL}},\quad
    R_\text{PVL} = 
    \begin{cases}
        R_\text{PVL,open}, \, P_\text{LV} < P_\text{Ao}\\
        R_\text{PVL,closed}, \, P_\text{LV} \ge P_\text{Ao}
    \end{cases}
\end{equation}
in which $P_\text{Ao}$ and $P_\text{LV}$ denote the downstream aortic and upstream ventricular pressures, respectively, and $R_\text{PVL}$ controls the magnitude of the leakage ($R_\text{PVL,closed}$ is a finite, large value, when the PVL pathway is closed).
Thus, this reduced-order correction accounts for the paravalvular component present in the experimental flow-rate measurement without requiring explicit simulation of the stent-deployment process and leak-gap formation.

We distinguish the experimentally measured flow-rate waveforms used during reduced-order calibration from the boundary fluxes computed by the fully coupled FSI model.
The quantities $Q_\text{LV}^\text{exp}$ and $Q_\text{Ao}^\text{exp}$ denote the measured upstream and downstream flow-rate waveforms used as calibration data.
By contrast, $Q_\text{in}$ and $Q_\text{out}$ denote the instantaneous volumetric fluxes through the inlet and outlet of the computational domain, respectively, and are determined by the FSI solution.
Using the outward unit normal $\n$, these fluxes are defined as
\begin{equation}
Q_\text{in} = -\int_{\Gamma_\text{in}}\u\cdot\n\,\Da, 
\qquad
Q_\text{out} = \int_{\Gamma_\text{out}}\u\cdot\n\,\Da,
\end{equation}
so that both are positive in the forward direction from the ventricular compartment toward the aortic compartment.
The experimentally measured waveforms are not imposed as boundary fluxes in the fully coupled FSI simulations.

Briefly, the resulting upstream model used in our simulations is given by
\begin{align}
\label{eq:PVL_correction_LV}
Q_\text{LV} &= Q_\text{in}+Q_\text{PVL},\\
\label{eq:PIV_upstream_model}
P_\text{LV} &= P_\text{pump}-Q_\text{LV}R_\text{up}.
\end{align}
All upstream flow rates are defined as positive in the forward direction from the ventricular compartment toward the aortic compartment.
With this convention, the negative value of $Q_\text{PVL}$ during regurgitation verifies the positive sign in Equation~\eqref{eq:PVL_correction_LV}.
Here, $R_\text{up}$ characterizes the upstream resistance, and $P_\text{pump}$ is the prescribed pump-pressure waveform constructed from the corresponding experimentally measured $P_\text{LV}^\text{exp}$ and $Q_\text{LV}^\text{exp}$ waveforms.
The quantity $Q_\text{in}$ is the computational inlet flux determined by the coupled FSI solution.
The total upstream flow $Q_\text{LV}$ is obtained by combining this computational flux with the auxiliary PVL flow according to Equation~\eqref{eq:PVL_correction_LV}.

For each candidate value of $R_\text{up}$, the pump-pressure waveform is constructed from the target upstream pressure and flow-rate waveforms as
\begin{equation}
P_\text{pump}(t;R_\text{up}) = P_\text{LV}^\star(t) + R_\text{up}Q_\text{LV}^\star(t),
\label{eq:pump_pressure_construction}
\end{equation}
in which the superscript $\star$ denotes the target waveform.
Combining Equations~\eqref{eq:PIV_upstream_model} and \eqref{eq:pump_pressure_construction} gives
\begin{equation}
Q_\text{LV}(t)-Q_\text{LV}^\star(t) = \frac{P_\text{LV}^\star(t)-P_\text{LV}(t)}{R_\text{up}}.
\end{equation}
Provided that the pressure difference remains bounded, the computed flow rate therefore approaches the target waveform as $R_\text{up}\rightarrow\infty$.
This limiting behavior arises because the constructed $P_\text{pump}$ depends explicitly on $R_\text{up}$ through the term $R_\text{up}Q_\text{LV}^\star$; it would not occur if $P_\text{pump}$ were held fixed while $R_\text{up}$ increased.
However, increasing $R_\text{up}$ also amplifies fluctuations in the target flow-rate waveform when constructing $P_\text{pump}$, which can introduce nonphysical pressure oscillations.
Therefore, $R_\text{up}$ is calibrated to reproduce the target flow rate as closely as possible while limiting oscillations in the derived pump-pressure waveform.

\begin{figure}[tt!]
	\begin{center}
		\includegraphics[scale = 0.225,trim={0 0 0 0},clip]{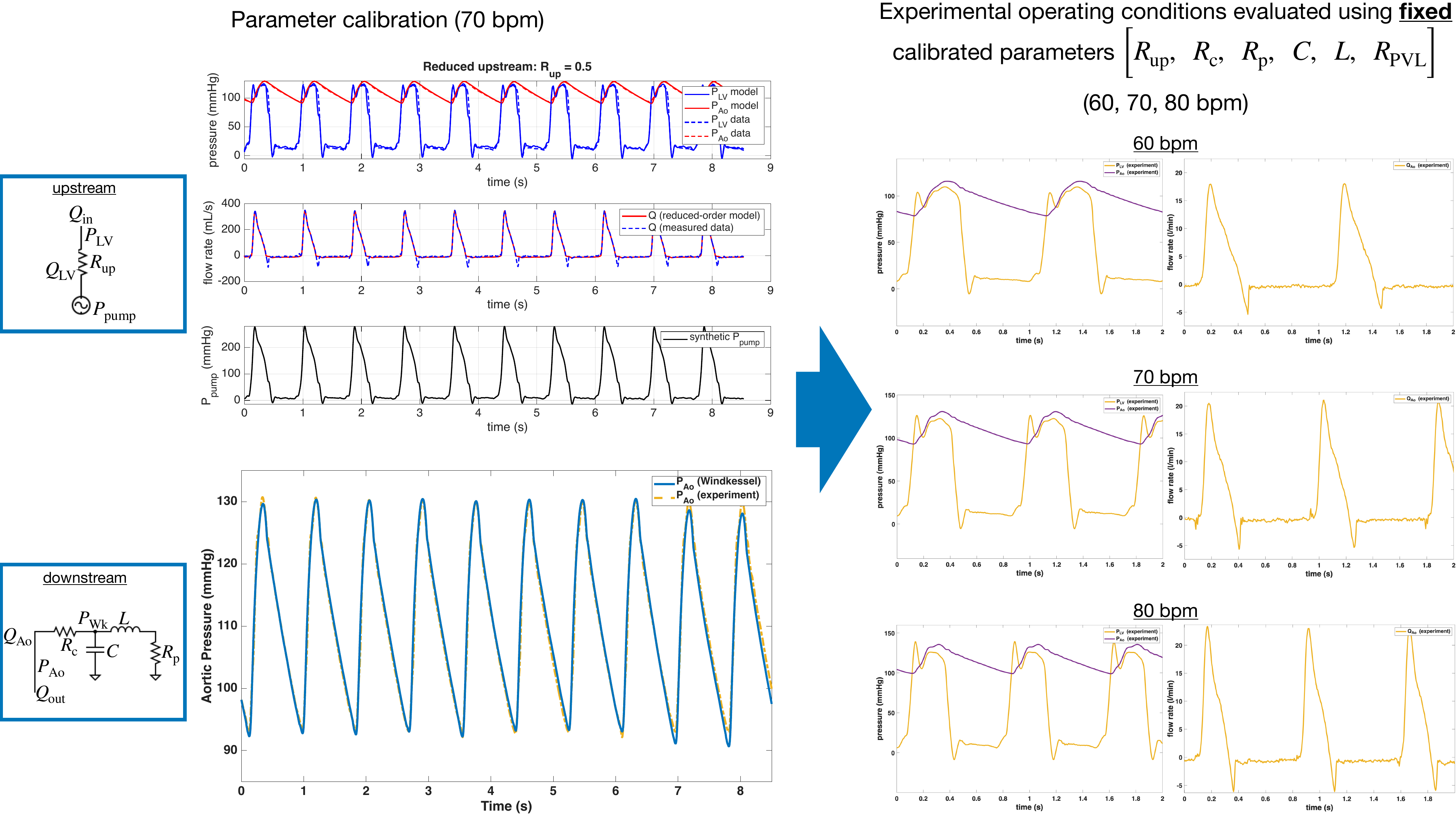}
		\caption{
        Boundary-model calibration and operating-condition evaluation workflow.
        The left panel summarizes the sequential calibration performed using the 70~bpm experimental data.
        In the calibration flow subplot, $Q$ denotes flow in $\text{mL}\,\text{s}^{-1}$, with positive values directed from the ventricular compartment toward the aortic compartment.
        The measured trace is obtained from the upstream flow probe, and an internal sign conversion reconciles the opposite upstream and transvalvular flow-variable orientations used in the reduced-order calibration code.
        The downstream RCLR Windkessel parameters and reduced-order valve parameters are identified first; these parameters are then held fixed while $R_\text{up}$ is calibrated and $P_\text{pump}$ is constructed.
        Following calibration, $R_\text{up}$, $R_\text{c}$, $R_\text{p}$, $C$, $L$, and $R_\text{PVL}$ remain fixed at all three pulse rates, whereas a condition-specific $P_\text{pump}$ waveform is constructed from the measured upstream pressure and flow rate at each pulse rate.
        The right panels show representative experimental pressure and flow-rate waveforms at 60, 70, and 80~bpm and do not represent FSI simulation results.
        The 60 and 80~bpm cases provide out-of-calibration evaluations of parameter transfer under condition-specific, measurement-informed upstream forcing.
        The corresponding FSI--experiment comparisons are presented in Figure~\ref{fig:complete_comparison}.
        }
		\label{fig:parameter_calibration}
	\end{center}
\end{figure}

The downstream loading conditions of the pulse duplicator are represented by a four-element RCLR Windkessel model,
\begin{align}
\label{eq:PVL_correction_Ao}
    Q_\text{Ao} &= Q_\text{out} + Q_\text{PVL},\\
\label{eq:windkessel_compliance}
C\frac{dP_\text{Wk}}{dt} &= Q_\text{Ao}-Q_\text{p},\\
\label{eq:windkessel_peripheral_branch}
L\frac{dQ_\text{p}}{dt} &= P_\text{Wk} - Q_\text{p}R_\text{p},\\
\label{eq:windkessel_aortic_pressure}
P_\text{Ao} &= P_\text{Wk} + Q_\text{Ao}R_\text{c}.
\end{align}
Here, $C$ is the compliance, $R_\text{c}$ is the characteristic resistance, $R_\text{p}$ is the peripheral resistance, and $L$ is the inertance.
The quantities $P_\text{Wk}$, $Q_\text{Ao}$, and $P_\text{Ao}$ are the Windkessel pressure, downstream boundary-model flow rate, and aortic pressure, respectively.
The quantity $Q_\text{out}$ is the computational outlet flux determined by the coupled FSI solution.
During reduced-order calibration, the experimentally measured waveform $Q_\text{Ao}^\text{exp}$ is supplied to the RCLR Windkessel model, but it is not imposed in the fully coupled FSI simulation.
The inertance $L$ is included to represent the inertial effects observed in the measured flow-rate data.

The upstream and downstream boundary models were calibrated using the experimental data acquired at 70~bpm, as summarized in Figure~\ref{fig:parameter_calibration}.
For the downstream model, the measured flow-rate waveform was supplied as input to the RCLR Windkessel model.
The parameters $R_\text{c}$, $R_\text{p}$, $C$, and $L$ were identified using the constrained nonlinear optimization routine \textsf{fmincon} in MATLAB by minimizing the discrepancy between the downstream pressure predicted by the Windkessel model and the corresponding experimental measurement.
For the downstream calibration, we quantified the model--data discrepancy using the $L^2$-norm of the difference between the predicted and measured $P_\text{Ao}$ waveforms.
The valve objective is the equally weighted sum of the squared residuals in $P_\text{LV}$, $P_\text{Ao}$, and $Q_\text{Ao}$, with each residual normalized by the standard deviation of the corresponding experimental waveform.

We evaluated a discrete set of candidate values of $R_\text{up}$ using the 70~bpm data.
For each candidate, we calculated NMSEs for $P_\text{LV}$, $P_\text{Ao}$, and $Q_\text{Ao}$ and defined the composite error as
\begin{equation}
E_\text{comp}
=
w_P\frac{E_{P_\text{LV}}+E_{P_\text{Ao}}}{2}
+
w_QE_{Q_\text{Ao}},
\qquad
w_P+w_Q=1.
\end{equation}
Using the primary weights $(w_P,w_Q)=(0.1,0.9)$, $R_\text{up}=0.5~\text{mmHg}\,\text{s}\,\text{mL}^{-1}$ minimized the composite error.
We assigned greater weight to the flow-rate error because the upstream model is designed primarily to reproduce the target flow-rate waveform as closely as possible while limiting induced oscillations in the pressure waveform.
The same value of $R_\text{up}$ was selected using $(w_P,w_Q)=(0.2,0.8)$, indicating that the selected value was unchanged between these two weight choices.

We determined $R_\text{PVL}$ separately using a manual bracket-and-bisection procedure with the 70~bpm data.
We first identified resistance values that underestimated and overestimated the measured diastolic regurgitant flow and then successively refined the interval until satisfactory agreement was obtained.
This procedure yielded $R_\text{PVL}=9.78~\text{mmHg}\,\text{s}\,\text{mL}^{-1}$, which remained fixed at all operating conditions.

Following this initial calibration, the downstream parameters remain fixed during the subsequent valve and upstream calibration stages.
We first identify the valve parameters using a reduced-order representation of the complete experimental loop and the constrained nonlinear optimization routine \textsf{fmincon} in MATLAB.
At this stage, the upstream subsystem follows the detailed pulse-duplicator model of Lee et al.~\cite{Lee2020pulse}, with its parameters fitted to the present 70~bpm experimental measurements and the measured left atrial pressure and pump-flow waveforms supplied as inputs.
The valve is represented by a direction-dependent pressure--flow relation containing effective viscous, convective, and unsteady inertial loss terms, following reduced-order valve formulations reported previously~\cite{Mynard2012,Keijsers2016,Ferlito2026}.
The governing equations, physical interpretations of the loss terms, and calibrated valve parameters are provided in Supplementary Information Section~\ref{supp:valve_upstream_calibration}.

After identifying the valve parameters, we replace the detailed upstream apparatus model with the minimal single-resistance upstream model used in the FSI simulations.
This representation retains dynamic pressure--flow coupling while avoiding the additional states and parameters of the complete pulse duplicator model. 
We then calibrate $R_{\text{up}}$ to reproduce the target flow rate waveform while minimizing induced pressure oscillations and subsequently construct the prescribed pump pressure waveform $P_{\text{pump}}(t)$.

We calibrate $R_{\text{PVL}}$ separately using the full FSI model at 70~bpm through a manual bracket-and-bisection procedure that matches the unresolved diastolic regurgitant flow.
The resulting $R_{\text{PVL}}$ and all other calibrated boundary-model parameter values remain fixed at all three pulse rates.
For each pulse rate, however, a condition-specific $P_{\text{pump}}(t)$ waveform is constructed from the corresponding measured upstream pressure and flow rate waveforms.

\subsection{Operating conditions and calibration protocol}
\label{sec:operating_conditions}

We performed experiments and fully coupled FSI simulations at pulse rates of 60, 70, and 80~bpm.
We designated 70~bpm as the boundary-parameter calibration condition and 60 and 80~bpm as out-of-calibration evaluation conditions.
We identified the reduced-order boundary-model parameters $R_\text{up}$, $R_\text{c}$, $R_\text{p}$, $C$, $L$, and $R_\text{PVL}$ exclusively using the experimental measurements obtained at 70~bpm.
After calibration, these parameters remained fixed and were applied without further adjustment at all three operating conditions.
Table~\ref{tab:boundary_model_parameters} reports the calibrated parameter values.

For each pulse rate, we constructed a condition-specific pump-pressure waveform, $P_\text{pump}(t)$, from the corresponding experimentally measured waveforms $P_\text{LV}^\text{exp}(t)$ and $Q_\text{LV}^\text{exp}(t)$.
Thus, the 60 and 80~bpm simulations involved no condition-specific recalibration of the boundary-model parameters but used measurement-informed upstream forcing constructed separately for each operating condition.
These cases therefore provide out-of-calibration evaluations of parameter transfer across pulse rates under condition-specific, measurement-informed upstream forcing.

Although the measured upstream waveforms informed the construction of $P_\text{pump}(t)$, neither $P_\text{LV}(t)$ nor $Q_\text{LV}(t)$ was imposed directly in the fully coupled FSI simulations.
Instead, both emerge from the interaction among the fluid, valve leaflets, and boundary models.
Consequently, their agreement with experiment assesses the consistency of the coupled simulation with the measurement-informed upstream forcing rather than providing independent validation of upstream pressure and flow.
The corresponding $P_\text{Ao}$ and PDVA measurements were not used to construct the forcing and therefore provide the principal out-of-calibration validation targets at 60 and 80~bpm.

\begin{table}[!t]
\centering
\caption{Summary of parameters and their values for the boundary models.}
\label{tab:boundary_model_parameters}
\begin{tabular}{@{}lcl@{}}
\hline
\textbf{Parameter} & \textbf{Value} & \textbf{Unit} \\
\hline
$R_\text{up}$  & 0.5   & $\text{mmHg}\,\text{s}\,\text{mL}^{-1}$ \\
$R_\text{p}$   & 2.629 & $\text{mmHg}\,\text{s}\,\text{mL}^{-1}$ \\
$R_\text{c}$   & 0.001 & $\text{mmHg}\,\text{s}\,\text{mL}^{-1}$ \\
$C$            & 0.882 & $\text{mL}\,\text{mmHg}^{-1}$ \\
$L$            & 0.145 & $\text{mmHg}\,\text{s}^{2}\,\text{mL}^{-1}$ \\
$R_\text{PVL, open}$ & 9.78  & $\text{mmHg}\,\text{s}\,\text{mL}^{-1}$ \\
$R_\text{PVL, closed}$ & 10000  & $\text{mmHg}\,\text{s}\,\text{mL}^{-1}$ \\
\hline
\end{tabular}
\end{table}

\subsection{Numerical discretizations and infrastructure}
Simulations were performed using the reconstructed geometry of the ViVitro PIV mount embedded in a computational domain with dimensions 10.1~cm $\times$ 5.05~cm $\times$ 5.05~cm.
We used an adaptively refined Cartesian grid to solve the Navier--Stokes equations, and we used an unstructured finite element (FE) mesh that conforms to the geometry of the structure~\cite{Griffith2017}.
The effective fine-grid resolution of the Cartesian grid is approximately 0.4~mm.
Structural models use second-order tetrahedral (10-point) elements for the TAVR leaflets and first-order tetrahedral (4-point) elements for the PIV mount, which were generated using \texttt{fTetWild}~\cite{Hu2020}.
The average grid spacings of the structural meshes are 0.66~mm for the valve leaflets and 0.495~mm for the remaining structures.
We used a piecewise-linear kernel for the aortic test section and the stent, as well as a three-point B-spline kernel for the valve leaflets as regularized delta functions.
The time-step size started at $\Delta t = 5.0 \times 10^{-6}$~s.

We solve the coupled FSI system using the immersed finite element--difference (IFED) method~\cite{Griffith2017,Griffith2020,Wells2023} implemented in the IBAMR software infrastructure, which is a distributed-memory parallel implementation of the IB method with adaptive mesh refinement (AMR)~\cite{Griffith2007, IBAMR}.
The governing equations and Eulerian--Lagrangian coupling formulation are described in Supplementary Information Section~\ref{supp:ifed_method}.
IBAMR uses SAMRAI~\cite{SAMRAIPaper} for Cartesian grid discretization management, libMesh~\cite{libMeshPaper} for finite element discretization management, and PETSc~\cite{petsc-web-page,petsc-user-ref,petsc-efficient} for linear solver infrastructure.

\subsection{Cycle averaging and quantitative comparison}
\label{sec:quantitative_comparison}

We assess the cycle-to-cycle consistency of the simulated waveforms using cycles 2--4.
For two successive cycles, $x_n(t)$ and $x_{n+1}(t)$, the waveforms are linearly interpolated onto a common time grid $\{t_j\}_{j=1}^{N}$ over their shared interval.
The successive-cycle NRMSE is defined as
\begin{equation}
\mathrm{NRMSE}_{n,n+1} = \frac{\left[\frac{1}{N}\sum_{j=1}^{N}\left(x_{n+1}(t_j)-x_n(t_j)\right)^2\right]^{1/2}}{\max_j x_{n+1}(t_j) - \min_j x_{n+1}(t_j)}\times 100\%.
\label{eq:successive_cycle_nrmse}
\end{equation}
Thus, the RMSE is normalized by the peak-to-peak range of the later cycle over the shared comparison interval.
We calculate this metric for cycles 2--3 and 3--4.

We also calculate an ensemble periodicity error using cycles 2--4.
The simulation ensemble mean is
\begin{equation}
\overline{x}(t_j) = \frac{1}{3}\sum_{k=2}^{4}x_k(t_j),
\label{eq:simulation_ensemble_mean}
\end{equation}
and the ensemble NRMSE is
\begin{equation}
\mathrm{NRMSE}_{\mathrm{ens}} = \frac{\left[\frac{1}{3N}\sum_{k=2}^{4}\sum_{j=1}^{N}\left(x_k(t_j)-\overline{x}(t_j)\right)^2\right]^{1/2}}{\max_j\overline{x}(t_j) - \min_j\overline{x}(t_j)}\times 100\%.
\label{eq:ensemble_cycle_nrmse}
\end{equation}
Thus, the ensemble mean is the reference waveform, and the squared deviations are pooled with equal weight across all three cycles and all time samples before normalization by the peak-to-peak range of the ensemble mean.
All calculations use only the time interval shared by the included waveforms.

\section{Results}
\label{sec:results}
We performed experiments and fully coupled FSI simulations at pulse rates of 60, 70, and 80~bpm.
For each operating condition, we used 10 consecutive experimental cycles to calculate ensemble-averaged pressure, flow rate, and PDVA waveforms and to characterize cycle-to-cycle experimental variability.
As described in Section~\ref{sec:operating_conditions}, we calibrated the reduced-order boundary-model parameters exclusively using the 70~bpm experimental data and subsequently held these parameters fixed at all three pulse rates.
We constructed a condition-specific, measurement-informed pump-pressure waveform at each pulse rate.
The 60 and 80~bpm cases therefore provide out-of-calibration evaluations of parameter transfer under measurement-informed upstream forcing.
We evaluated the coupled model by comparing the simulated and experimental pressure and flow-rate waveforms, leaflet kinematics, and PDVA.

\subsection{Flow-field evolution}
\label{subsec:flow_field_results}

\begin{figure}[tt!]
	\begin{center}
		\includegraphics[scale = 0.24,trim={0 0 0 0},clip]{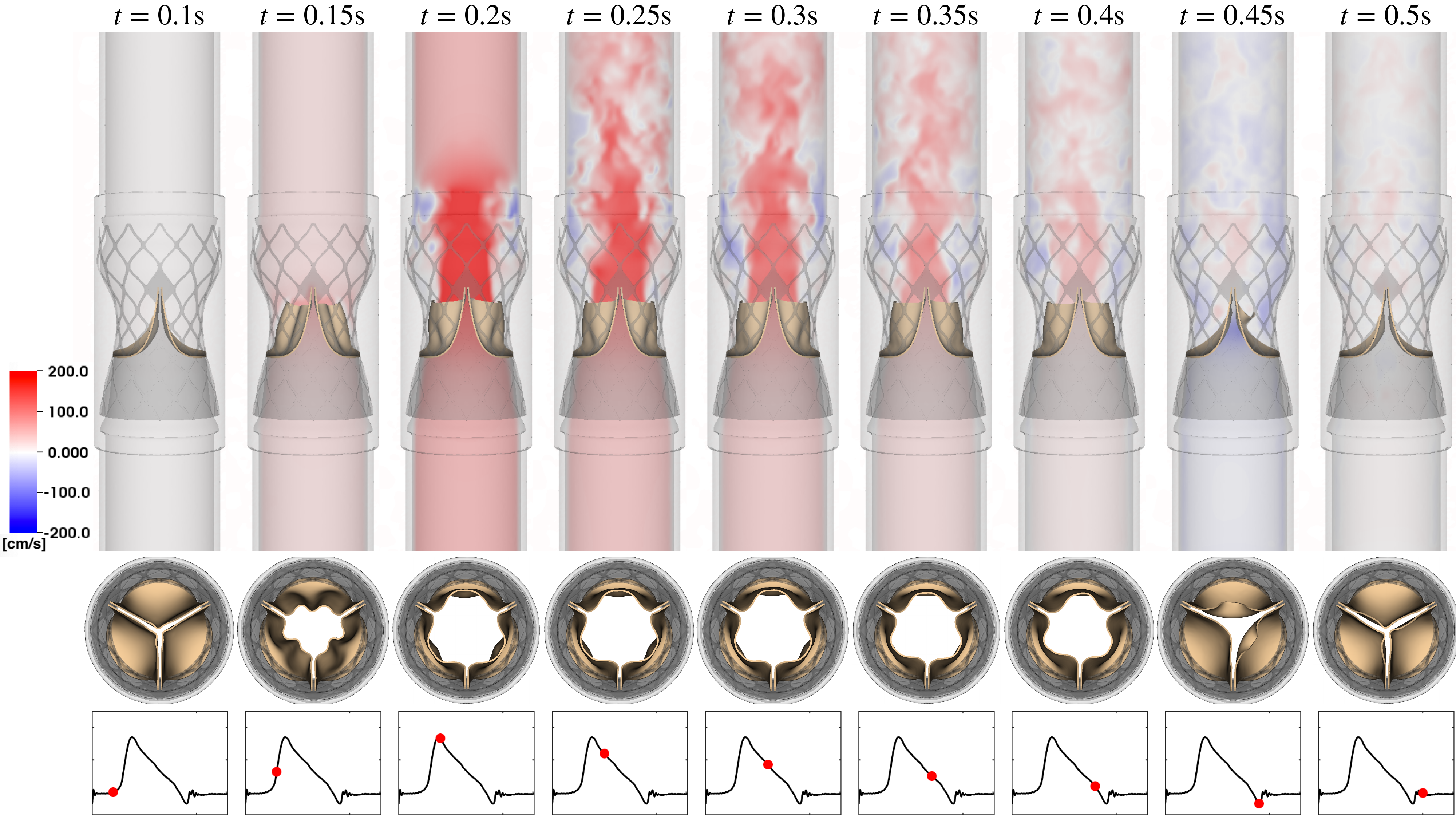}
		\caption{
        Representative evolution of the simulated valve and flow field at 60~bpm.
        The upper row shows a longitudinal section through the computational model at selected times during the cardiac cycle.
        Color indicates the axial velocity component, with red denoting flow in the forward direction and blue denoting reverse flow.
        The middle row shows a top view of the valve, illustrating leaflet kinematics at the selected times.
        The lower row identifies each displayed time on the corresponding aortic flow waveform.
        }
		\label{fig:U_z_60bpm}
	\end{center}
\end{figure}

Figure~\ref{fig:U_z_60bpm} illustrates the evolution of the simulated valve and surrounding flow field over a representative cardiac cycle at 60~bpm.
A brief region of reverse axial velocity develops during valve closure, coinciding with the negative portion of the flow-rate waveform.
These results demonstrate that the FSI model resolves the coupled progression of valve opening, forward-flow acceleration, flow deceleration, brief flow reversal, and valve closure.

\subsection{Effect of paravalvular leakage pathway}
\label{subsec:pvl_results}
\begin{figure}[tt!]
	\begin{center}
		\includegraphics[scale = 0.24,trim={0 0 0 0},clip]{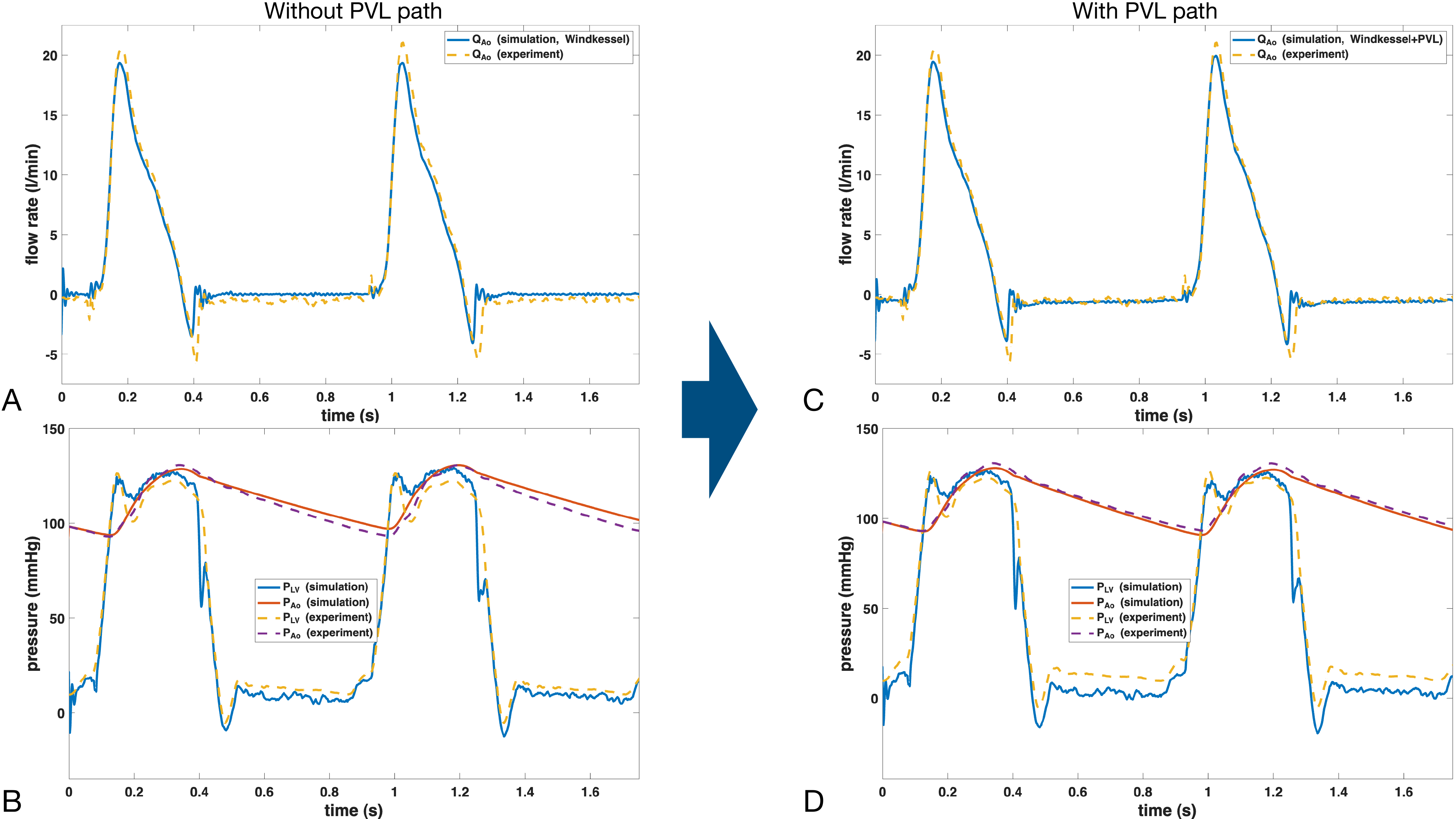}
		\caption{
        Effect of the reduced-order paravalvular leak (PVL) pathway on the predicted flow rate and pressure waveforms at 70~bpm.
        \textbf{A--B.} Comparison without the PVL pathway.
        The uncorrected simulation reproduces the primary forward-flow waveform but does not account for the complete regurgitant flow measured experimentally after valve closure.
        \textbf{C--D.} Comparison after including the PVL pathway.
        The total flow supplied to the reduced-order circulation is computed by combining the flow through the simulated test section with the estimated PVL flow.
        This correction improves agreement with the measured regurgitant flow and produces the corresponding downstream-pressure response.
        }
		\label{fig:comparison_before_after_pvl}
	\end{center}
\end{figure}

Figure~\ref{fig:comparison_before_after_pvl} illustrates the effect of including the reduced-order PVL pathway in the comparison between the experimental and simulated hemodynamics.
The experimental flow-rate measurement includes flow through both the valve orifice and paravalvular gaps between the implanted device and the surrounding test section.
Because the computational geometry does not explicitly represent these gaps, the uncorrected FSI flow rate includes only the flow through the modeled test section and does not reproduce the full experimentally measured regurgitant flow.
Consequently, the model without the PVL pathway reproduces the principal systolic flow waveform but underestimates the negative flow measured after valve closure (Figure~\ref{fig:comparison_before_after_pvl}A), which then leads to increased downstream pressure (Figure~\ref{fig:comparison_before_after_pvl}B).

The lumped PVL pathway estimates the unresolved leakage flow from the instantaneous upstream-to-downstream pressure difference and combines it with the flow through the computational domain.
Including the PVL improves agreement with the experimentally measured regurgitant flow during diastole while preserving the overall systolic waveform (Figure~\ref{fig:comparison_before_after_pvl}C).
The PVL correction also supplies the downstream Windkessel model with the total flow entering the experimental downstream circulation, producing the appropriate downstream-pressure response (Figure~\ref{fig:comparison_before_after_pvl}D).

\subsection{Pressure and flow-rate comparison}
\label{subsec:hemodynamic_results}

\begin{figure}[tt!]
	\begin{center}
		\includegraphics[scale = 0.425]{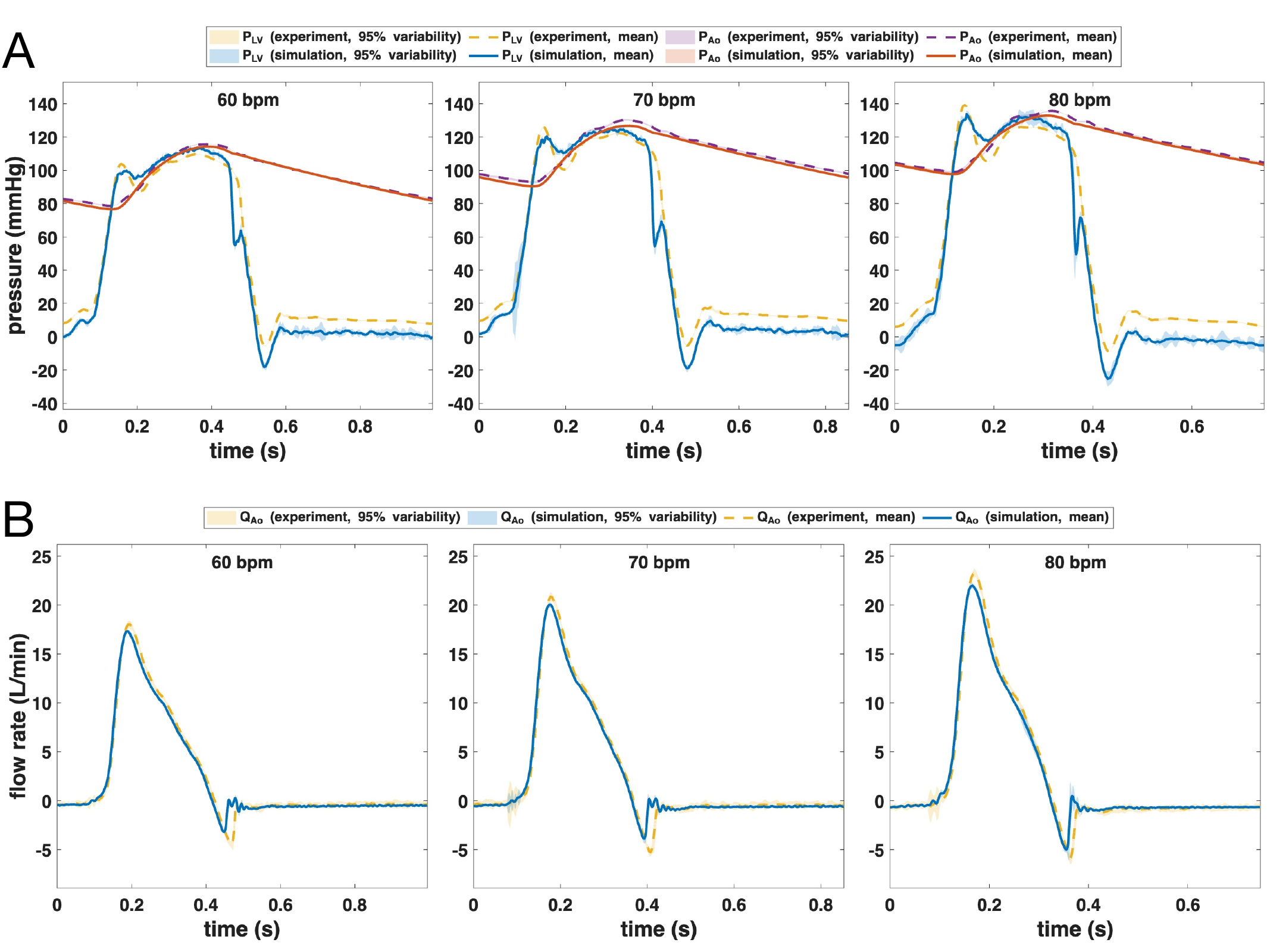}
		\caption{
        Comparison of experimental and simulated pressure and flow waveforms at 60, 70, and 80~bpm for the selected ($R_\text{up}=0.5$ $\text{mmHg}\,\text{s}\,\text{mL}^{-1}$) boundary-model case.
        \textbf{A.}
        Left ventricular pressure, $P_\text{LV}$, and aortic pressure, $P_\text{Ao}$.
        \textbf{B.}
        Aortic volumetric flow rate, $Q_\text{Ao}$.
        Experimental dashed curves represent cycle-averaged waveforms calculated from 10 consecutive cardiac cycles, and the corresponding shaded regions represent approximate pointwise 95\% cycle-to-cycle variability bands defined as the experimental mean $\pm\, 1.96$ standard deviations.
        The simulated shaded regions represent the pointwise mean $\pm1.96$ standard deviations across cycles 2--4 and are included as a descriptive visualization of cycle-to-cycle dispersion.
        They are not confidence intervals for the simulated mean.
        Given the limited sample of three simulated cycles, the displayed dispersion bands are not interpreted as statistically estimated 95\% population intervals.
        All waveforms are plotted against time measured from the beginning of the analyzed cardiac cycle.
        The close agreement among simulation cycles 2--4 indicates that the pressure and flow solutions approached a periodic state; comparison with the experimental ensemble characterizes agreement with the measured hemodynamics.
        }
		\label{fig:complete_comparison}
	\end{center}
\end{figure}

Figure~\ref{fig:complete_comparison} compares the simulated and experimental pressure and flow-rate waveforms at 60, 70, and 80~bpm.
The 60 and 80~bpm simulations use the boundary-model parameter values calibrated at 70~bpm without further adjustment.
However, the prescribed pump-pressure waveform is constructed separately at each pulse rate using the corresponding measured upstream pressure and flow-rate waveforms.
Accordingly, agreement in $P_\text{LV}$ and flow rate primarily demonstrates that the coupled simulation remains consistent with the measurement-informed upstream forcing rather than providing independent validation of these quantities.
By contrast, the measured $P_\text{Ao}$ and PDVA waveforms at 60 and 80~bpm are not used to construct the forcing or recalibrate the model parameters.
Their comparison with the corresponding simulation results therefore provides the principal out-of-calibration validation evidence for the downstream hemodynamic response and valve dynamics.
The simulated flow-rate waveforms closely reproduce the timing and magnitude of the experimental forward-flow phase at all three operating conditions.
The model also captures the rapid flow deceleration and brief regurgitant-flow phase associated with valve closure.
The most noticeable localized flow-rate differences occur during and immediately after closure, when the waveforms change rapidly and exhibit short-duration oscillations.

The downstream aortic pressure, $P_\text{Ao}$, agrees closely with the experimental measurement throughout the cardiac cycle at all three pulse rates.
The reduced-order PVL pathway accounts for the paravalvular component contained in the experimental flow-rate measurement, allowing the flow supplied to the downstream Windkessel model to reproduce the experimentally measured total aortic flow.
The resulting downstream pressure waveforms reproduce both the systolic pressure levels and the gradual pressure decrease during diastole.

The simulated upstream ventricular pressure, $P_\text{LV}$, captures the principal systolic features of the experimental waveform, including the rapid pressure increase, the systolic plateau, and the pressure decrease during valve closure.
A modest difference remains during diastole, when the simulated $P_\text{LV}$ generally remains below the experimental mean.
This difference is associated with the auxiliary PVL pathway, which modifies the coupling between the upstream and downstream reduced-order boundary models if $P_\text{Ao}>P_\text{LV}$.
The influence of this reduced-order PVL representation on the upstream pressure is considered further in Section~\ref{sec:discussion}.

The widths of the simulated variability bands are small over most of the cardiac cycle, indicating close agreement among simulation cycles 2--4.
The greatest simulated cycle-to-cycle differences are localized near rapid pressure transitions and valve closure.
These results indicate that the simulated pressure and flow rate solutions approach a periodic state over the analyzed cycles.
We assess cycle-to-cycle consistency quantitatively using the successive-cycle and ensemble NRMSEs defined in Equations~\eqref{eq:successive_cycle_nrmse} and~\eqref{eq:ensemble_cycle_nrmse}.
Across $P_\text{LV}$, $P_\text{Ao}$, $Q_\text{Ao}$, and PDVA, the maximum successive-cycle NRMSE is 1.46\%, 2.06\%, and 2.34\% at 60, 70, and 80~bpm, respectively.
The ensemble NRMSE across cycles 2--4 remains below 1.16\% for every signal and operating condition.
These small errors demonstrate close consistency among cycles 2--4 and support the use of their mean for comparison with experiment.
Because only three cycles are analyzed, these results support an approximately periodic response over the analyzed cycles rather than formally establishing asymptotic convergence to a periodic solution.

\subsection{Leaflet kinematics and projected dynamic valve area}
\label{subsec:pdva_results}

\begin{figure}[tt!]
	\begin{center}
		\includegraphics[scale = 0.24,trim={0 0 0 0},clip]{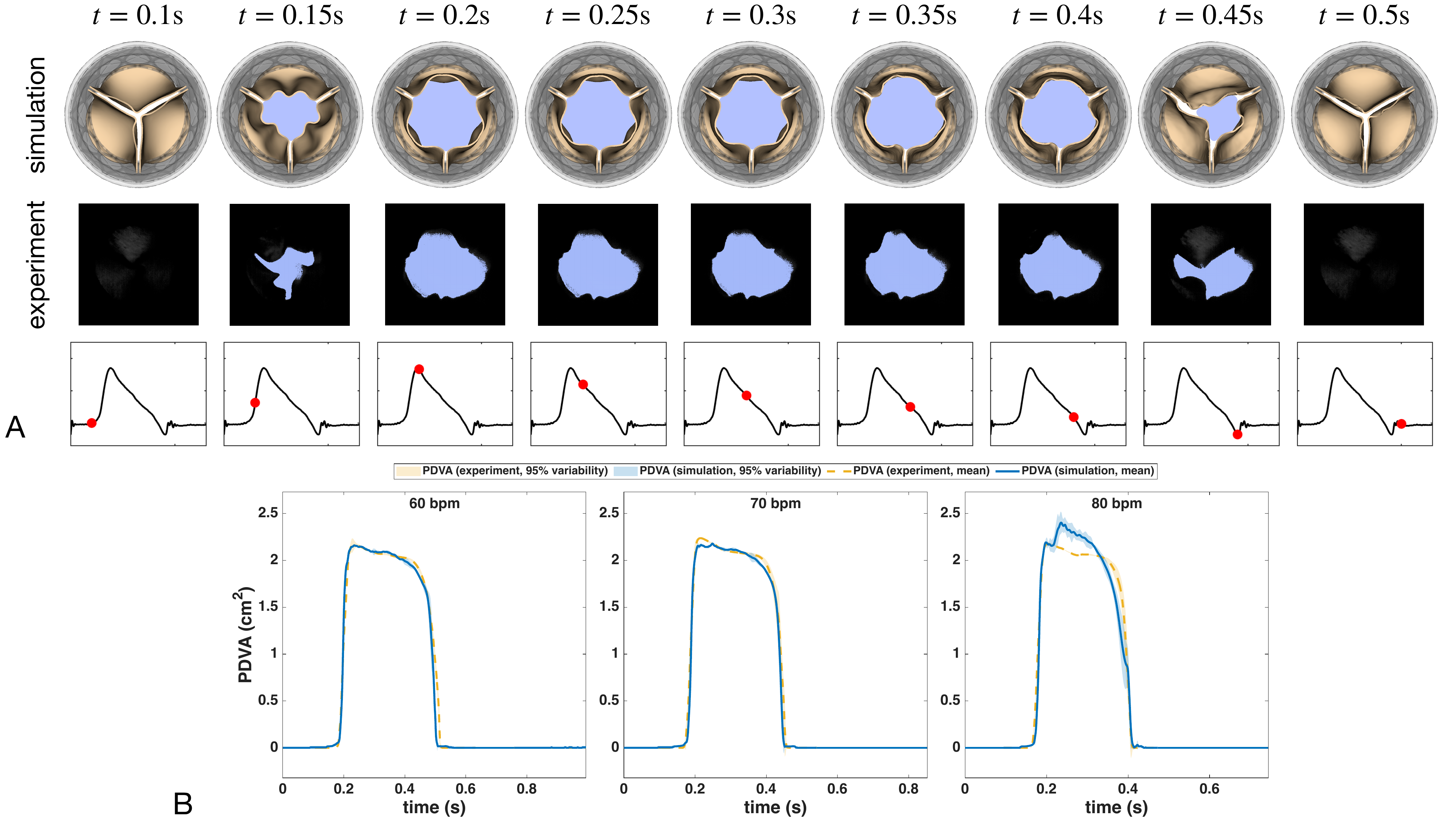}
		\caption{
        Comparison of simulated and experimental valve kinematics and projected dynamic valve area (PDVA).
        \textbf{A.}
        Representative en face valve configurations from the simulation are shown in the upper row, and phase-aligned high-speed experimental images are shown in the middle row.
        The highlighted regions identify the projected open orifice used to determine PDVA.
        The experimental frames were manually phase-aligned to the simulated sequence using the onset of valve opening; therefore, the times shown above the columns denote the nominal simulation times and should not be interpreted as exact experimental acquisition times.
        The lower row identifies the corresponding locations on the aortic flow waveform.
        \textbf{B.}
        Experimental and simulated PDVA waveforms at 60, 70, and 80~bpm.
        Experimental curves represent the mean PDVA calculated from 10 consecutive cardiac cycles, with shaded regions indicating approximate pointwise 95\% cycle-to-cycle variability bands defined as the experimental mean $\pm\, 1.96$ standard deviations.
        Simulated curves represent the mean of simulation cycles 2--4, with shaded regions indicating the simulated mean $\pm\, 1.96$ standard deviations across those cycles.
        The simulated bands describe cycle-to-cycle dispersion and are not confidence intervals for the mean or statistically estimated 95\% population intervals.
        }
		\label{fig:PDVA}
	\end{center}
\end{figure}

Figure~\ref{fig:PDVA} compares the simulated and experimental leaflet kinematics and PDVA.
The same pixel-to-length scale determined from the 70~bpm images was used for the 60 and 80~bpm PDVA measurements without additional calibration.
This procedure is appropriate because the valve, test section, camera position, image resolution, and optical configuration remained unchanged, whereas pulse rate was the only intentionally varied experimental condition.
Using a single spatial calibration also prevents condition-specific rescaling from artificially improving agreement between the experimental and simulated PDVA.
The representative en face view images (Figure~\ref{fig:PDVA}A) show that the simulation reproduces the principal stages of the experimentally observed valve motion, including the onset of opening, expansion to a fully open configuration, maintenance of the open orifice during systole, and subsequent closure.
Differences in the detailed leaflet-edge configurations remain visible, particularly during the rapidly changing opening and closing phases.
These differences are consistent with the use of an idealized computational geometry with three nominally identical leaflets.

The PDVA waveforms provide a quantitative comparison of the simulated and experimental valve kinematics (Figure~\ref{fig:PDVA}B).
At 60 and 70~bpm, the simulation closely reproduces the timing of valve opening, the maximum projected area, the systolic evolution of the open area, and the timing of valve closure.
At 80~bpm, the simulation reproduces the overall opening and closing sequence but exhibits a somewhat larger peak PDVA and greater variability during the early open phase.
Nevertheless, the simulated and experimental waveforms remain similar in their overall magnitude and temporal evolution across all three operating conditions.

The simulated PDVA variability bands are calculated from cycles 2--4 and remain narrow over most of the cardiac cycle.
This consistency indicates that the simulated leaflet kinematics, in addition to the bulk pressure and flow-rate waveforms, approach a periodic state.
The agreement at 60 and 80~bpm is particularly notable because the PDVA measurements at these operating conditions are not used either to calibrate the reduced-order boundary-model parameters or to construct the condition-specific upstream forcing.

Table~\ref{tab:nrmse_results} summarizes the agreement between the cycle average of simulation cycles 2--4 and the cycle average of 10 experimental cycles.
The simulated flow rate includes the reduced-order PVL correction.

\begin{table}[ht!]
\centering
\caption{Simulation versus experiment normalized root-mean-square error (NRMSE) (\%) for the cycle-averaged waveforms.
The simulated mean uses cycles 2--4, and the experimental mean uses 10 cycles.
Each RMSE is normalized by the peak-to-peak range of the corresponding experimental mean over the shared comparison interval.}
\label{tab:nrmse_results}
\begin{tabular}{lrrrr}
\hline
Pulse rate & $P_\text{LV}$ & $P_\text{Ao}$ & $Q_\text{Ao}$ & PDVA \\
\hline
60~bpm & 7.64 & 3.50 & 3.08 & 4.12 \\
70~bpm & 7.64 & 6.21 & 3.03 & 3.07 \\
80~bpm & 8.07 & 5.05 & 2.88 & 4.70 \\
\hline
\end{tabular}
\end{table}

For the cycle-mean simulation--experiment comparisons, the aortic flow NRMSE ranges from 2.88\% to 3.08\%, whereas the PDVA NRMSE ranges from 3.07\% to 4.70\%.
Upstream ventricular pressure has the largest NRMSE, ranging from 7.64\% to 8.07\%, whereas downstream aortic pressure NRMSE ranges from 3.50\% to 6.21\%.

The full-cycle PDVA NRMSE combines discrepancies during leaflet motion with the extended interval during which both the experimental and simulated PDVA are approximately zero.
We therefore perform supplementary comparisons over the active-phase and open-configuration intervals.
The active phase includes samples for which either cycle-mean waveform exceeds 5\% of the experimental peak.
The open configuration includes samples for which the experimental cycle-mean waveform is at least 90\% of its peak.
For all three PDVA comparisons, the RMSE evaluated over the specified interval is normalized by the peak-to-peak range of the experimental cycle-mean PDVA over the full shared comparison interval, rather than by its range within the active-phase or open-configuration interval.

\begin{figure}[ht!]
\centering
\includegraphics[width=\textwidth]{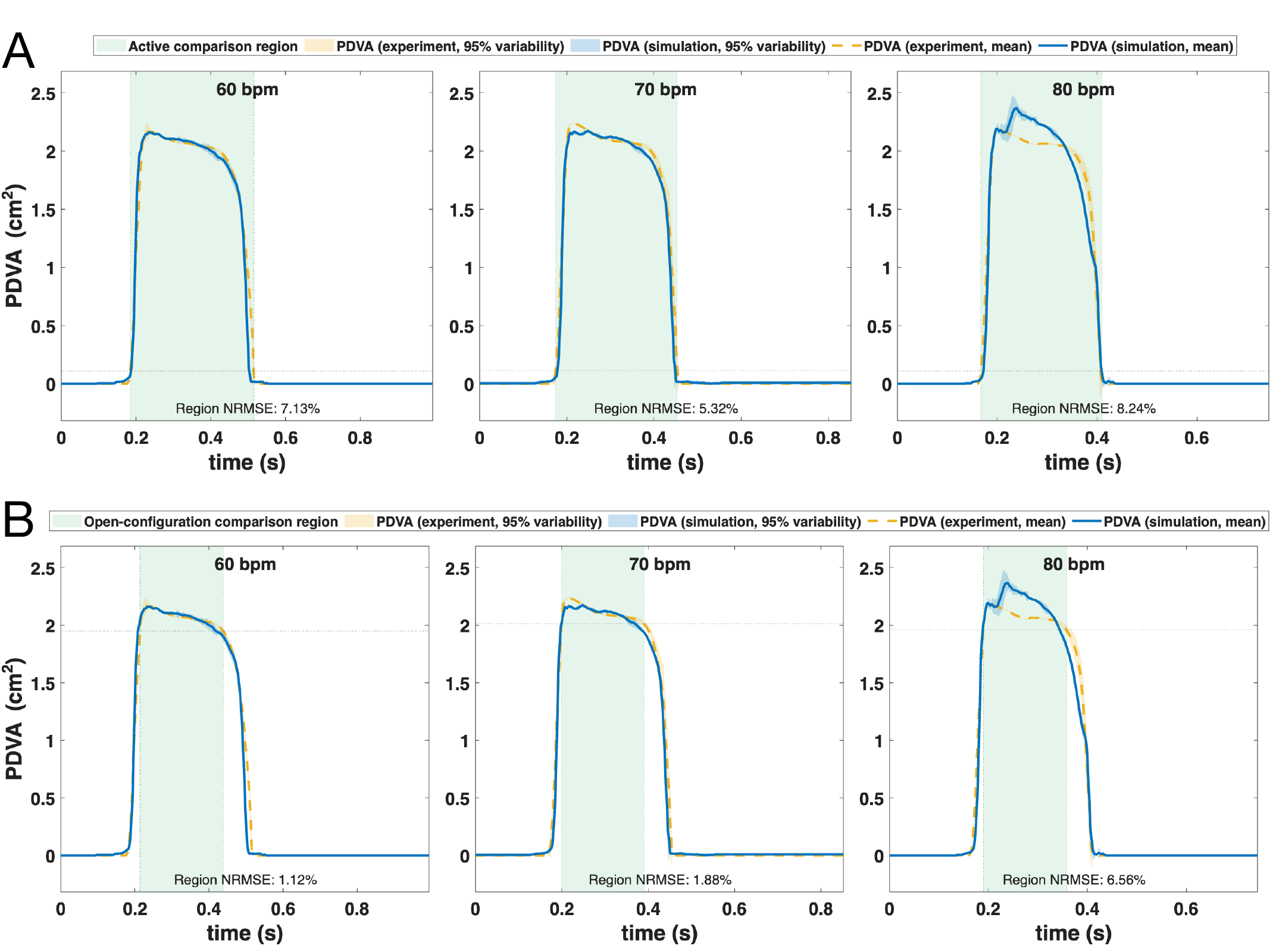}
\caption{
    \textbf{A.}
    Active-phase PDVA comparison regions at 60, 70, and 80~bpm.
    Green shading identifies samples for which either cycle-averaged waveform exceeds 5\% of the experimental peak, and the horizontal dotted line indicates this threshold.
    \textbf{B.}
    Open-configuration PDVA comparison regions.
    Green shading identifies samples for which the experimental cycle average is at least 90\% of its peak, and the horizontal dotted line indicates this threshold.
    Yellow dashed and blue solid curves denote the experimental and simulated cycle averages, respectively.
    Shaded bands show the corresponding mean $\pm\, 1.96$ standard deviations, calculated from 10 experimental cycles and simulation cycles 2--4, respectively.
    The simulated bands describe cycle-to-cycle dispersion and are not confidence intervals for the mean or statistically estimated 95\% population intervals.
}
\label{fig:NRMSE_PDVA}
\end{figure}

Figure~\ref{fig:NRMSE_PDVA} shows the exact time regions included in these supplementary calculations (Table~\ref{tab:pdva_configuration_results}).
Green vertical shading identifies the evaluated samples, and dotted vertical lines delimit each contiguous comparison region.
The active-phase region includes opening, sustained opening, and closure, whereas the open-configuration region isolates the near-maximum experimental opening.

The open-configuration NRMSE is 1.12\% at 60~bpm but increases to 6.56\% at 80~bpm, despite the relatively similar full-cycle NRMSEs of 4.12\% and 4.70\%, respectively.
Let $A_\text{exp}(t)$ and $A_\text{sim}(t)$ denote the experimental and simulated cycle-mean PDVA waveforms, respectively.
The peak-area error is defined as
\begin{equation}
E_\text{peak} = \frac{\max\limits_t A_\text{sim}(t)-\max\limits_t A_\text{exp}(t)}{\max\limits_t A_\text{exp}(t)} \times 100\%.
\end{equation}
A positive value indicates overprediction of the maximum PDVA, whereas a negative value indicates underprediction.
The corresponding peak-area errors are $-0.29\%$ at 60~bpm and $+8.72\%$ at 80~bpm.

To define the closure timing, $t_\text{closure,exp}$ and $t_\text{closure,sim}$ are taken as the first descending crossings of one-half of the respective waveform maximum after the global peak.
The crossing times are obtained by linear interpolation between adjacent samples.
The closure delay is defined as
\begin{equation}
\Delta t_\text{closure} = 1000\left(t_\text{closure,sim}-t_\text{closure,exp}\right) \ \text{ms}.
\end{equation}
A positive closure delay indicates that simulated closure occurs later than experimental closure, whereas a negative value indicates earlier simulated closure.
The half-peak crossing provides a reproducible measure of closure timing and does not represent the instant of complete leaflet coaptation.
These metrics distinguish the close sustained-opening agreement at 60~bpm from the larger open-area overprediction at 80~bpm.

\begin{table}[ht!]
\centering
\small
\caption{Supplementary PDVA agreement metrics between the simulation mean of cycles 2--4 and the experimental mean of 10 cycles.
Positive peak error denotes overprediction of the maximum PDVA, and positive closure delay denotes later simulated closure.}
\label{tab:pdva_configuration_results}
\begin{tabular}{lrrrrr}
\hline
Pulse rate & Full NRMSE & Active NRMSE & Open NRMSE & Peak error & Closure delay \\
& (\%) & (\%) & (\%) & (\%) & (ms) \\
\hline
60~bpm & 4.12 & 7.13 & 1.12 & $-0.29$ & $-5.36$ \\
70~bpm & 3.07 & 5.32 & 1.88 & $-2.84$ & $-3.03$ \\
80~bpm & 4.70 & 8.24 & 6.56 & $+8.72$ & $-7.86$ \\
\hline
\end{tabular}
\end{table}

\section{Discussion}
\label{sec:discussion}

This study builds upon our previously developed FSI model of a bovine pericardial bioprosthetic heart valve in an experimental pulse duplicator~\cite{Lee2020pulse,Lee2021flutter}.
We extend this framework to a TAVR configuration and compare simulated and experimental hemodynamics and leaflet kinematics under pulsatile operating conditions.
The circulation parameters were calibrated using reduced-order models; the PVL resistance was calibrated separately using the 70~bpm FSI simulation.
Although the measured upstream pressure and flow-rate waveforms were not imposed directly, they were used to construct the condition-specific pump-pressure waveform prescribed at each pulse rate.

Figure~\ref{fig:complete_comparison} shows close agreement between the simulated and experimental flow-rate waveforms at 60, 70, and 80~bpm.
The simulations reproduce the magnitude and timing of the principal forward-flow features across all three operating conditions.
The most noticeable localized differences occur during valve closure, when rapid flow deceleration and brief reversal make the response particularly sensitive to leaflet geometry, material behavior, contact, and fluid--structure coupling.
The simulated downstream aortic pressure, $P_\text{Ao}$, also agrees closely with the experimental waveform, whereas a modest discrepancy remains in the upstream ventricular pressure, $P_\text{LV}$, particularly during diastole. This discrepancy is expected because we imposed the upstream boundary model to act as closely as possible as a flow source.

The experimental flow-rate measurement contains a paravalvular component that is absent from the directly resolved FSI flow.
Although the computational geometry includes an idealized stent in its prescribed deployed configuration, the model does not simulate stent deployment, contact, or apposition against the test-section wall and therefore does not resolve the resulting paravalvular gaps.
We account for this unresolved flow using an auxiliary diode-like pathway with a conducting-state resistance, $R_{\text{PVL}}$, calibrated at 70~bpm and held fixed across all operating conditions, that connects the upstream and downstream boundary models.
The combined model captures the net paravalvular component of the experimental flow-rate measurement and provides the total flow entering the downstream Windkessel model.
Including this contribution preserves the experimentally observed negative flow during diastole and contributes to the close agreement between the simulated and experimental $P_\text{Ao}$ waveforms.

The remaining diastolic discrepancy in $P_\text{LV}$ appears to arise from the interaction between the lumped PVL pathway and the minimal upstream model.
During diastole, when $P_{\text{Ao}}>P_{\text{LV}}$, the PVL pathway produces retrograde flow according to Equation~\eqref{eq:pvl_eqn}.
At the same time, $P_\text{LV}$ is constrained by the prescribed pump-pressure waveform and the single upstream resistance, whereas $P_\text{Ao}$ is determined by the PVL-corrected flow supplied to the downstream Windkessel model.
The coupled system must therefore maintain the pressure difference required to reproduce the negative PVL flow while remaining consistent with the corrected downstream pressure.
Because the minimal upstream model does not represent all of the compliant, inertial, and transient dynamics of the physical upstream system, it may not reproduce the measured diastolic $P_\text{LV}$ exactly under these coupled constraints.
The discrepancy therefore reflects a tradeoff associated with reproducing the measured PVL flow and downstream pressure while using a simplified upstream representation that acts approximately as a flow source.

Agreement between the experiment and FSI simulation at 70~bpm is not a direct or trivial consequence of calibrating the reduced-order boundary models at this operating condition.
During calibration, we supplied the experimentally measured flow-rate waveform as an input to the reduced-order system and selected the boundary-model parameters to reproduce the measured pressure response.
In the fully coupled FSI simulation, however, neither the experimental flow-rate waveform nor its corresponding pressure response is imposed directly.
Instead, both quantities emerge from the interaction among the calibrated boundary models, fluid dynamics, and deforming valve leaflets.
Agreement at 70~bpm therefore indicates that the calibrated boundary models remain suitable for use with the detailed FSI model and that the modeled leaflet mechanics and fluid--structure coupling produce a consistent hemodynamic response.

Because the measured $P_\text{LV}$ and $Q_\text{LV}$ waveforms at each pulse rate are used to construct the corresponding prescribed $P_\text{pump}$ waveform, agreement in upstream pressure and flow should not be interpreted as independent validation of these quantities.
Neither waveform is imposed directly in the fully coupled FSI simulation; instead, both emerge from the interaction among the upstream boundary model, fluid, valve leaflets, and downstream loading.
Their agreement is therefore not automatic, but it represents consistency with measurement-informed upstream forcing rather than an independent prediction.

The principal out-of-calibration validation evidence is provided by quantities that are not used to construct the condition-specific forcing, particularly $P_\text{Ao}$ and PDVA at 60 and 80~bpm.
These comparisons evaluate whether the parameters calibrated at 70~bpm, together with the coupled valve model, reproduce the downstream hemodynamic response and leaflet dynamics at other pulse rates without parameter recalibration.
The results therefore provide evidence that the calibrated parameter values transfer across operating conditions, conditional on experimentally informed upstream forcing.

The experimental upstream-pressure waveform exhibits increasingly prominent transient peaks as the pulse rate increases.
These features limit the ability of the reduced-order upstream model to reproduce every detail of the measured waveform.
One possible explanation is that increasing the pulse rate produces fluid accumulation and transient pressurization within a portion of the upstream compliance chamber that is not explicitly represented by the reduced-order model.
The observed pressure buildup near the viewing window supports the presence of experimental system dynamics that are not fully captured by the current lumped-parameter representation.
Additional measurements and modeling are needed to determine the physical origin of these transient pressure features.

The PDVA comparison provides a complementary assessment of the model by evaluating leaflet kinematics more directly than the bulk pressure and flow-rate waveforms.
The simulation reproduces the principal features of the experimental PDVA waveforms, including the onset of opening, the rapid increase in projected area, the approximately sustained open configuration during systole, and the subsequent decrease in area during valve closure.
The narrow simulated PDVA variability bands across cycles 2--4 further indicate that the leaflet motion approaches a periodic state together with the pressure and flow-rate waveforms.
The agreement in the magnitude and timing of PDVA at 60 and 70~bpm indicates that the model captures the dominant opening and closing behavior of the valve.
The close agreement in both the magnitude and timing of PDVA at 60~bpm is particularly noteworthy because the 60~bpm PDVA measurements are not used either to calibrate the boundary-model parameters or to construct the condition-specific upstream forcing.
The ability of the model to reproduce this independently evaluated output indicates that the coupled model captures the dominant relationships among boundary loading, transvalvular hemodynamics, and leaflet motion.

At 80~bpm, the simulation reproduces the overall temporal evolution of PDVA but predicts a somewhat larger peak area and greater cycle-to-cycle variability during the early open phase.
Despite these differences, the agreement remains encouraging because the 80~bpm PDVA measurements are not used either to calibrate the boundary-model parameters or to construct the condition-specific upstream forcing.
Thus, the PDVA comparisons at 60 and 80~bpm provide out-of-calibration validation evidence for valve dynamics under condition-specific, measurement-informed upstream forcing.

The difference between the 60 and 80~bpm validation results may be related to the pulse-rate dependence of the transient ``bump'' observed in the experimental upstream-pressure waveform.
Relative to the 70~bpm calibration condition, this feature becomes smaller at 60~bpm but more pronounced at 80~bpm.
The reduced-order upstream boundary model does not explicitly represent the physical mechanism responsible for this transient pressure feature.
The parameters calibrated at 70~bpm may therefore transfer more readily to 60~bpm because the unmodeled contribution is weaker at that rate.
At 80~bpm, the larger pressure transient produces a greater departure from the upstream behavior represented during calibration and may contribute to the observed differences in PDVA.
In particular, differences in the transient transvalvular pressure may alter the fluid loading on the leaflets and thereby affect the predicted maximum opening area and its cycle-to-cycle variability.
This interpretation remains a potential explanation rather than a definitive causal conclusion, and additional measurements are needed to identify the origin of the pulse-rate-dependent pressure feature.

The agreement across operating conditions is also notable because the leaflet constitutive parameters are obtained from the measurements reported by Murdock et al.~\cite{Murdock2018} rather than from mechanical testing of the specific valve specimen used in the pulse-duplicator experiments.
Considering the transfer of the boundary-model parameters from 70~bpm, the pulse-rate-dependent upstream-pressure transient, and the use of non-specimen-specific leaflet properties, the close PDVA agreement at 60~bpm and the good overall agreement at 80~bpm provide encouraging evidence of predictive capability for valve dynamics under measurement-informed upstream forcing and without condition-specific parameter recalibration.

Several limitations of the valve model may also contribute to the remaining differences between simulation and experiment.
First, the computational model uses an idealized geometry with three identical leaflets, whereas the experimental valve may exhibit geometric and mechanical asymmetry.
Future work will incorporate specimen-specific geometries reconstructed from micro-CT.

Second, the leaflet constitutive parameters are based on measurements reported by Murdock et al.~\cite{Murdock2018} rather than on biaxial testing of the experimental valve specimen.
Specimen-specific testing is not possible without removing and destroying the implanted leaflets.
Differences in tissue properties may therefore contribute to discrepancies in PDVA, flutter, and valve-closure dynamics.

Third, the model does not explicitly simulate stent deployment or the formation of gaps between the prosthesis and the surrounding wall.
The lumped $R_{\text{PVL}}$ pathway captures the net paravalvular leakage measured experimentally but does not resolve the spatial structure of the leakage jets.

Finally, the present validation focuses on bulk pressure, flow rate, and PDVA.
Future work will compare phase-averaged experimental and simulated velocity fields to assess the spatially resolved pulsatile flow.

The scope of this validation is limited to the modeled valve specimen, pulse-duplicator configuration, tested pulse rates, and evaluated quantities of interest.
These results do not by themselves establish predictive credibility for other valve specimens, patient-specific conditions, in vivo applications, or model outputs not evaluated in this study.
This study provides experimental bench validation evidence for selected quantities of interest and is not intended to constitute a complete risk-informed credibility assessment under ASME V\&V~40~\cite{ASMEVV40} or FDA computational modeling guidance~\cite{FDACredibility}.

Despite these limitations, agreement at 70~bpm provides a consistency assessment at the calibration condition.
The 60 and 80~bpm comparisons provide conditional validation evidence for downstream aortic pressure and PDVA under condition-specific, measurement-informed upstream forcing.
The 60 and 80~bpm results further demonstrate that the boundary-model parameters calibrated at 70~bpm transfer across pulse rates without condition-specific parameter recalibration, although the upstream forcing is informed by measurements acquired at each operating condition.
This framework may support future investigations of intravalvular and paravalvular regurgitation, reduced leaflet mobility, subclinical leaflet thrombosis, and pannus formation~\cite{DeMarchena2015}.

\section*{Acknowledgments}
This work was supported by National Institutes of Health (NIH) Awards R01HL157631, R03HL182166, and U01HL143336. Additional research support was provided by National Science Foundation (NSF) Award OAC 1931516. Computations were performed using facilities provided by Old Dominion University through Research Computing. The Wahab cluster is supported in part by NSF Award CNS-1828593, ``MRI Acquisition: A Reconfigurable Computing Infrastructure Enabling Interdisciplinary and Collaborative Research in Hampton Roads.'' We also thank D. Adalsteinsson for support with ImageTank.

This manuscript is the result of funding in whole or in part by the National Institutes of Health (NIH). It is subject to the NIH Public Access Policy. Through acceptance of this federal funding, NIH has been given a right to make this manuscript publicly available in PubMed Central upon the Official Date of Publication, as defined by NIH.

\section*{Conflict of Interest}
No conflict of interest.

\section*{Author Contributions}
J.H.L. and B.E.G. contributed to conceptualization. J.H.L., P.A., and A.D.N. contributed to data curation. J.H.L., P.A., and M.S. contributed to formal analysis. B.E.G. acquired funding. J.H.L. and P.A. conducted the investigation. J.H.L., P.A., A.D.N., R.H., and W.K. contributed to methodology. J.H.L. and B.E.G. were responsible for project administration and provided resources. B.E.G. provided supervision. J.H.L., P.A., and M.S. contributed to validation. J.H.L. and P.A. prepared the visualizations. J.H.L., M.S., P.A., and A.D.N. wrote the original draft. J.H.L., M.S., and B.E.G. reviewed and edited the manuscript.

\section*{Data Availability}
The data generated and/or analyzed during the current study are available from the corresponding author upon reasonable request.

\clearpage
\bibliography{reference}


\clearpage
\section*{Supplementary Information}
\renewcommand\thefigure{S\arabic{figure}}
\renewcommand\thetable{S\arabic{table}}
\renewcommand\theequation{S\arabic{equation}}
\renewcommand\theHequation{S\arabic{equation}}
\setcounter{equation}{0}
\setcounter{figure}{0}
\setcounter{table}{0}

\appendix

\section{Calibration of the reduced-order valve and upstream models}
\label{supp:valve_upstream_calibration}

During valve-parameter identification, the reduced-order valve is embedded in a representation of the complete experimental loop.
The upstream subsystem follows the detailed pulse-duplicator model of Lee et al.~\cite{Lee2020pulse}, with its parameters fitted to the present 70~bpm experimental measurements.
The measured left atrial pressure $P_{\text{LA}}^{\text{exp}}(t)$ and pump flow $Q_{\text{pump}}^{\text{exp}}(t)$ waveforms are prescribed as inputs, and the downstream subsystem uses the RCLR parameters calibrated previously from the 70~bpm measurements.
Table~\ref{tab:detailed_upstream_parameters} reports the fitted upstream parameters used during valve calibration.

\begin{table}[!t]
\centering
\caption{Parameters of the detailed upstream model used during
reduced-order valve calibration.}
\label{tab:detailed_upstream_parameters}
\begin{tabular}{@{}lcl@{}}
\hline
\textbf{Parameter} & \textbf{Value} & \textbf{Unit} \\
\hline
$R_{\text{out}}$       & 0.002216 & $\text{mmHg}\,\text{s}\,\text{mL}^{-1}$ \\
$R_{\text{VIA}}$       & 0.15     & $\text{mmHg}\,\text{s}\,\text{mL}^{-1}$ \\
$R_{\text{MV,open}}$   & 0.003467 & $\text{mmHg}\,\text{s}\,\text{mL}^{-1}$ \\
$R_{\text{MV,closed}}$ & $10^{8}$ & $\text{mmHg}\,\text{s}\,\text{mL}^{-1}$ \\
$C_{\text{VIA1}}$      & 0.2070   & $\text{mL}\,\text{mmHg}^{-1}$ \\
$C_{\text{VIA2}}$      & 0.003164 & $\text{mL}\,\text{mmHg}^{-1}$ \\
\hline
\end{tabular}
\end{table}

During boundary-model calibration, the valve is represented using a direction-dependent reduced-order pressure--flow relation~\cite{Mynard2012,Keijsers2016,Ferlito2026},
\begin{equation}
P_\text{LV}-P_\text{Ao} = L_\text{AV}\frac{\text{d}Q_\text{AV}}{\text{d}t} +
\begin{cases}
R_\text{AV,f}Q_\text{AV} +B_\text{AV,f}Q_\text{AV} \lvert Q_\text{AV}\rvert, & Q_\text{AV}\geq 0,\\
R_\text{AV,b}Q_\text{AV} +B_\text{AV,b}Q_\text{AV} \lvert Q_\text{AV}\rvert, & Q_\text{AV}<0.
\end{cases}
\label{eq:directional_valve_model}
\end{equation}
Here, $Q_\text{AV}$ is the transvalvular flow rate, and the subscripts $\text{f}$ and $\text{b}$ denote forward and reverse flow, respectively.
The terms $R_\text{AV,f}Q_\text{AV}$ and $R_\text{AV,b}Q_\text{AV}$ represent direction-dependent effective viscous losses.
The terms $B_\text{AV,f}Q_\text{AV}\lvert Q_\text{AV}\rvert$ and $B_\text{AV,b}Q_\text{AV}\lvert Q_\text{AV}\rvert$ represent direction-dependent losses associated with convective acceleration and dynamic pressure.
$L_\text{AV}\d{Q_\text{AV}}{t}$ represents the pressure difference associated with unsteady fluid inertia~\cite{Mynard2012,Keijsers2016,Ferlito2026}.
Separate loss coefficients are permitted for forward and reverse flow to represent the strongly direction-dependent hydraulic behavior of the valve.

This formulation is based on the pressure--flow relationship used in the versatile valve model of Mynard et al.~\cite{Mynard2012} and its subsequent applications to reduced-order valve dynamics~\cite{Keijsers2016,Ferlito2026}.
Unlike the complete Mynard model, the present formulation does not evolve a valve-opening state or calculate the loss and inertance coefficients from a time-dependent effective orifice area.
Instead, the coefficients are treated as effective parameters and are identified directly from the experimental hemodynamic measurements.

The valve parameters are identified using the constrained nonlinear optimization routine \textsf{fmincon} in MATLAB by minimizing the discrepancies between the modeled and measured upstream pressure, downstream pressure, and transvalvular flow rate waveforms.
The resulting parameters used during the subsequent upstream calibration are reported in Table~\ref{tab:reduced_valve_model_parameters}.

\begin{table}[!t]
\centering
\caption{Summary of parameters and their values for the reduced-order valve model.}
\label{tab:reduced_valve_model_parameters}
\begin{tabular}{@{}lcl@{}}
\hline
\textbf{Parameter} & \textbf{Value} & \textbf{Unit} \\
\hline
$R_\text{AV,f}$ & 0.0082 &
$\text{mmHg}\,\text{s}\,\text{mL}^{-1}$ \\
$B_\text{AV,f}$ & 0 &
$\text{mmHg}\,\text{s}^{2}\,\text{mL}^{-2}$ \\
$L_\text{AV}$ & 0.003 &
$\text{mmHg}\,\text{s}^{2}\,\text{mL}^{-1}$ \\
$R_\text{AV,b}$ & 7.5557 &
$\text{mmHg}\,\text{s}\,\text{mL}^{-1}$ \\
$B_\text{AV,b}$ & 0 &
$\text{mmHg}\,\text{s}^{2}\,\text{mL}^{-2}$ \\
\hline
\end{tabular}
\end{table}
Because the adopted quadratic coefficients are zero, the calibrated model used in the second stage reduces to
\begin{equation}
P_\text{LV}-P_\text{Ao} = L_\text{AV}\frac{\text{d}Q_\text{AV}}{\text{d}t} +
\begin{cases}
R_\text{AV,f}Q_\text{AV}, & Q_\text{AV}\geq 0,\\
R_\text{AV,b}Q_\text{AV}, & Q_\text{AV}<0.
\end{cases}
\label{eq:calibrated_directional_valve}
\end{equation}
Thus, its final calibrated form retains only the direction-dependent linear loss and unsteady inertance terms.

The calibrated valve parameters were subsequently held fixed while $R_\text{up}$ was identified using a reduced representation of the upstream--valve--downstream system.
For each candidate value of $R_\text{up}$, the corresponding pump-pressure waveform was constructed as
\begin{equation}
P_\text{pump}(t) = P_\text{LV,exp}(t) + R_\text{up}Q_\text{LV,exp}(t),
\label{eq:pump_pressure_construction_supp}
\end{equation}
subject to the adopted flow-direction convention.
We then selected $R_\text{up}$ to reproduce the measured flow-rate and upstream-pressure waveforms while limiting oscillations in the derived pump-pressure waveform.
The resulting $R_\text{up}$ and $P_\text{pump}(t)$ define the upstream loading condition used in the fully coupled FSI simulations.
The PVL resistance is not included in the reduced-order valve or upstream calibration.
Instead, $R_{\text{PVL}}$ is calibrated separately using the 70~bpm full FSI simulation through a manual bracket-and-bisection procedure that matches the unresolved diastolic regurgitant flow.
The calibrated value is subsequently held fixed at all three pulse rates.

\section{Immersed finite element--difference method}
\label{supp:ifed_method}
The immersed finite element--difference (IFED) method~\cite{Griffith2017,Griffith2020,Wells2023} is used to simulate our FSI model. The method uses an Eulerian description of the momentum, viscosity, and incompressibility of the coupled fluid--structure system and a Lagrangian description of the deformations, stresses, and resultant forces generated by the immersed structure.
In particular, we model the immersed structure as a viscoelastic solid in which the viscous stresses are small compared to the elastic stresses, as in previous work by us and others~\cite{Zhang2004, Liu2006, Zhang2007, Wang2010, Wang2012, Griffith2017, Griffith2020, Wells2023}.
The coupling between Eulerian and Lagrangian variables is mediated by integral transforms with delta-function kernels.
\begin{figure}[tt!]
	\begin{center}
		\includegraphics[scale = 0.35]{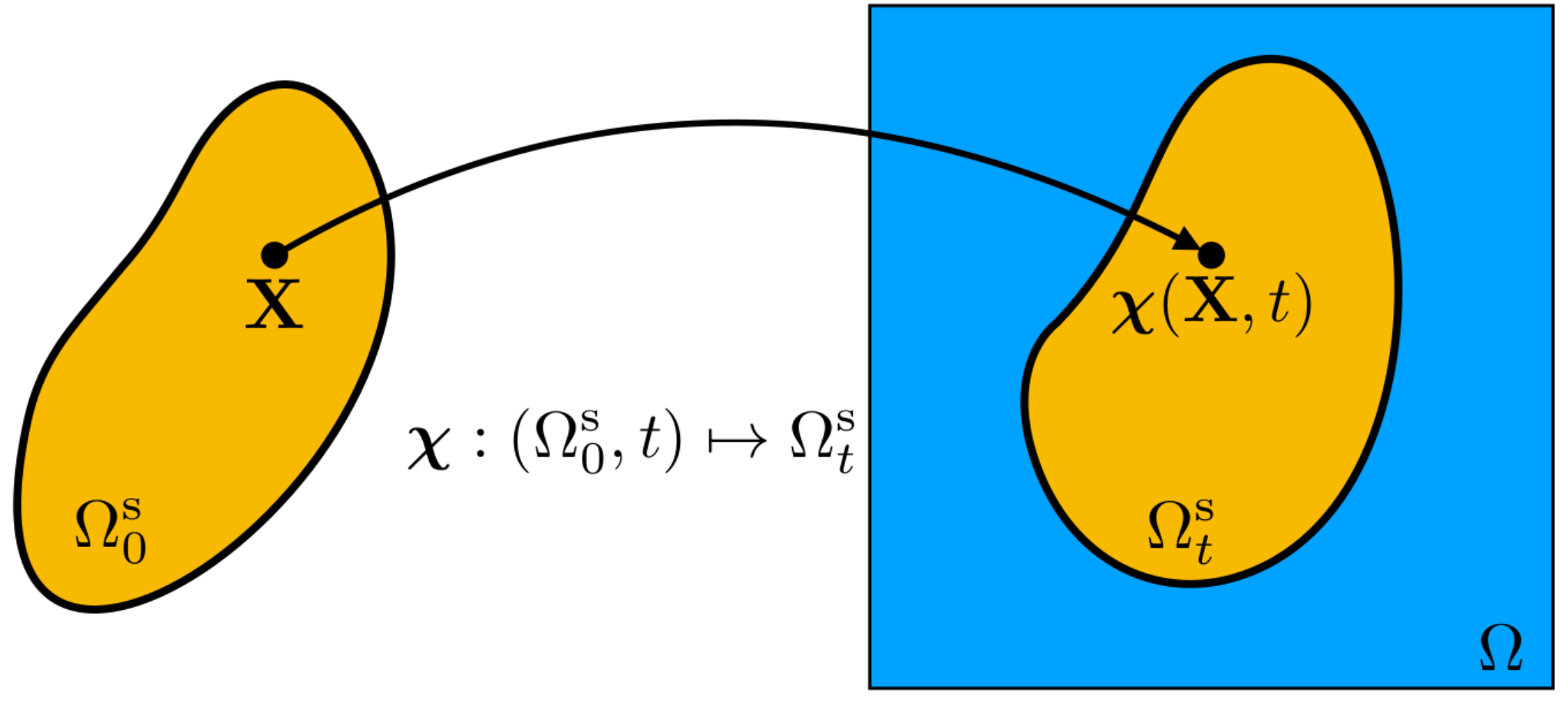}
		\caption{A schematic of the immersed boundary formulation. In the continuous IB formulation, we consider a fixed three-dimensional Eulerian computational domain $\Omega$ that is divided into the time-dependent solid ($\Omega^\text{s}_t$) and fluid ($\Omega^\text{f}_t$) subdomains, so that $\Omega = \Omega^\text{s}_t \cup \Omega^\text{f}_t$. $\vchi(\X,t) \in \Omega^\text{s}_t$ is the physical position of material point $\X$ at time $t$.}
		\label{fig:IBFE}
	\end{center}
\end{figure}
In the continuous IB formulation, we consider a fixed three-dimensional Eulerian computational domain $\Omega$ that is divided into the time-dependent solid ($\Omega^\text{s}_t$) and fluid ($\Omega^\text{f}_t$) subdomains, so that $\Omega = \Omega^\text{s}_t \cup \Omega^\text{f}_t$.
$\x = (x_1, x_2, x_3) \in \Omega$ are physical coordinates, $\X = (X_1, X_2, X_3) \in \Omega^\text{s}_0$ are reference coordinates attached to the structure, $\N(\X)$ is the outward unit normal to the reference configuration of the solid region at material point $\X\in \partial\Omega^\text{s}_0$, and $\vchi(\X,t) \in \Omega^\text{s}_t$ is the physical position of material point $\X$ at time $t$.
$\rho$ and $\mu$ are the mass density and viscosity, $\u(\x,t)$ and $p(\x,t)$ are the Eulerian velocity and pressure fields, $\f(\x,t)$ is the Eulerian elastic force density, $\PP(\X,t)$ is the first Piola--Kirchhoff elastic stress of the elastic body, and $\delta(\x) = \prod_{i=1}^{3}\delta(x_{i})$ is the three-dimensional Dirac delta function.
The IB form of the equations of motion for the fluid--structure system is:
 \begin{align}
    \label{eq:momentum}  \rho\frac{{\text D}\u}{{\text D}t}(\x,t) &= - \grad p(\x,t) + \mu \grad^2 \u(\x,t) + \f(\x,t), \\
    \label{eq:continuity}     \grad \cdot \u(\x,t) &= 0, \\
    \label{eq:fsiconstraint} \f(\x,t) &= \int_{\Omega^s_0} \grad_{\X} \cdot \PP(\X,t) \, \delta(\x - \vchi(\X,t)) \, \DX \\
    	     & \ \ \ \mbox{} - \int_{\p {\Omega^s_0}} \PP(\X,t) \, \N(\X) \, \delta(\x - \vchi(\X,t)) \, \DA ,\nonumber  \\
    \label{eq:noslip}           \D{\vchi}{t}(\X,t) &= \int_\Omega \u(\x,t) \, \delta(\x - \vchi(\X,t)) \, \Dx = \u(\vchi(\X,t),t).
  \end{align}
Notice that because $\p \vchi/\p t(\X,t) = \u(\vchi(\X,t),t)$ and $\grad \cdot \u(\x,t) = 0$, the immersed structure is described in this framework as an exactly incompressible material.

\end{document}